\documentclass[11pt]{article}
\usepackage{amsmath,amssymb,graphicx}
\usepackage{hyperref}
\usepackage[nosort]{cite}
\usepackage{multicol,color,longtable}

\usepackage{cite}
\usepackage{amsmath,amsfonts,amssymb,calrsfs}
\usepackage[small,bf,hang]{caption}
\usepackage{slashed}
\usepackage{latexsym,epsfig}

\definecolor{darkred}{rgb}{0.65,0.15,0}
\definecolor{darkgreen}{rgb}{.05,.5,.05}
\hypersetup{pdfborder={0 0 0},colorlinks=true,urlcolor=darkred,citecolor=blue,linkcolor=darkred,linktocpage=true}

\usepackage{mathrsfs}
\usepackage{dsfont}
\usepackage[Symbol]{upgreek}

\DeclareMathAlphabet{\mathpzc}{OT1}{pzc}{m}{it}

\newcommand{\nn}{\nonumber}

\newcommand{\reals}{\mathbb{R}}

\newcommand{\cN}{{\mathcal{N}}}

\newcommand{\cL}{{\mathcal{L}}}

\newcommand{\be}{\begin{align}}
\newcommand{\ee}{\end{align}}

\newcommand{\dL}{{\mathsf{d}}}
\newcommand{\dK}{{\mathsf{K}}}

\newcommand{\lb}{\left[}
\newcommand{\rb}{\right]}
\newcommand{\id}{\mathds{1}}

\newcommand{\mf}[1]{{\mathfrak{#1}}}

\makeatletter

\@addtoreset{equation}{section}
\makeatother

\newcommand{\Umat}[1]{\overset{#1}{U}{}}
\newcommand{\Kmat}[1]{\overset{#1}{C}{}}
\newcommand{\Jmat}[1]{\overset{#1}{\underline{J}}{}}

\newcommand{\Tr}{{\mathrm{Tr}}}

\newcommand{\bea}{\begin{eqnarray}}
\newcommand{\eea}{\end{eqnarray}}

\def\fg{{\mathfrak g}}

\def\fe{{\mathfrak e}}

\def\fsl{\mathfrak{sl}}

\def\dcox{g^\vee}
\def\Vir{\hbox{Vir}}

\def\*{\partial}
\def\punkt{\,\,.}

\def\komma{\,\,,}

\def\={\!=\!}

\def\eg{{\it e.g.}}

\def\ie{{\it i.e.}}

\def\bra#1{{\langle#1|}}
\def\ket#1{{|#1\rangle}}

\def\adj{\hbox{adj}}

\def\ZZ{{\mathds Z}}

\def\LL{{\cal L}}
\def\leftbr{[\![}
\def\rightbr{]\!]}

\begin{document}

{\flushright {CPHT-RR050.082017}\\[15mm]}

\begin{center}
  {\LARGE \bf \sc  Generalised diffeomorphisms for E$_9$}
    \\[15mm]

{\large
Guillaume Bossard${}^{1}$, Martin Cederwall${}^{2}$, Axel Kleinschmidt${}^{3,4}$, \\[1ex]Jakob Palmkvist${}^2$, Henning Samtleben${}^5$}

\vspace{15mm}
${}^1${\it Centre de Physique Th\'eorique, Ecole Polytechnique, CNRS\\
Universit\'e Paris-Saclay FR-91128 Palaiseau cedex, France}
\vskip 1 em
${}^2${\it Division for Theoretical Physics, Department of Physics,\\Chalmers University of Technology
 SE-412 96 Gothenburg, Sweden}
\vskip 1 em
${}^3${\it Max-Planck-Institut f\"{u}r Gravitationsphysik (Albert-Einstein-Institut)\\
Am M\"{u}hlenberg 1, DE-14476 Potsdam, Germany}
\vskip 1 em
${}^4${\it International Solvay Institutes\\
ULB-Campus Plaine CP231, BE-1050 Brussels, Belgium}
\vskip 1 em
${}^5${\it Univ Lyon, Ens de Lyon, Univ Claude Bernard, CNRS,\\
Laboratoire de Physique, FR-69342 Lyon, France}

\end{center}

\vfill 

\begin{center} 
  \textbf{Abstract}
\end{center} 
\begin{quote}

We construct generalised diffeomorphisms for E$_9$ exceptional field
  theory. The transformations, which like in the E$_8$ case contain
  constrained local transformations, close when acting on fields.
  This is the first example of a generalised diffeomorphism algebra based on an
  infinite-dimensional Lie algebra and an infinite-dimensional
  coordinate module.
  As a byproduct, we give a simple generic expression for the
  invariant tensors used in any extended geometry.
  We perform a generalised Scherk--Schwarz reduction and verify that our
  transformations reproduce the structure of gauged supergravity in
  two dimensions.
  The results are valid also for other affine algebras.

\end{quote} 
\vfill

\newpage

\tableofcontents


\section{Introduction}


Exceptional symmetries are one of the deepest features of ungauged
maximal supergravity, and symmetry groups of split real form E$_n(\mathds{R})$
have been established for $n\leq 9$, corresponding to supergravity in
$D=11-n$ space-time
dimensions \cite{Cremmer:1978ds,Julia:1982gx,Nicolai:1987kz,Cremmer:1997ct,Cremmer:1998px}. These   
symmetries are not only important for constructing gauged supergravity
models with interesting vacuum
structures, 
but also play an important role for understanding the string
theory effective action that conjecturally exhibits a discrete
U-duality symmetry E$_n(\mathds{Z})$~\cite{Hull:1994ys}, at least for
$n\leq 7$. 

Many papers have been devoted to understanding the origin of the
E$_n(\mathds{R})$ hidden symmetries, 
and recently there has been considerable progress on ``geometrising'' the
E$_n(\mathds{R})$ symmetries for $n\leq 8$. This geometrisation
requires first of all constructing an extended geometry that has
E$_n(\reals)$ symmetry and then, secondly, constructing a model based
on this so called exceptional geometry.
A crucial role in both steps is played
by a constraint on the geometry called the (strong) section constraint
that is necessary for defining a consistent algebra of generalised
diffeomorphisms and for making sure that the resulting
exceptional field theory reduces consistently to just standard
supergravity for a particular choice of exceptional geometric
background.
All these steps have been carried out for finite-dimensional
E$_n(\mathds{R})$ for $n\leq 8$ in a series of papers
\cite{Hull:2007zu,Pacheco:2008ps,Hillmann:2009pp,Berman:2010is,Berman:2011pe,Coimbra:2011ky,Coimbra:2012af,Berman:2012vc,Park:2013gaj,Cederwall:2013naa,Cederwall:2013oaa,Aldazabal:2013mya,Hohm:2013pua,Blair:2013gqa,Hohm:2013vpa,Hohm:2013uia,Hohm:2014fxa,Cederwall:2015ica}.
More generally, one can consider generalised
Scherk--Schwarz reductions of these theories
\cite{Berman:2012uy,Musaev:2013rq,Godazgar:2013oba,Lee:2014mla,Hohm:2014qga}
to obtain gauged supergravity
theories.  

In the present paper, we will begin the construction of E$_9$
exceptional field theory, where E$_9$ denotes the affine extension of
the largest finite-dimensional exceptional Lie group E$_8$. This
infinite-dimensional group
is known to be a symmetry of
two-dimensional maximal ungauged
supergravity~\cite{Julia:1982gx,Nicolai:1987kz}, and gaugings of this
symmetry have been considered in~\cite{Samtleben:2007an}. The first
step in the construction of E$_9$ exceptional field theory is
to establish a consistent gauge algebra of generalised diffeomorphisms
similar to~\cite{Berman:2012vc,Hohm:2014fxa,Cederwall:2015ica}.
This requires identifying an
appropriate set of coordinates that transform under E$_9$ together
with section constraints. They allow the definition of a
generalised Lie derivative that forms a closed algebraic structure. A
construction of a model based on the E$_9$ exceptional geometry will
be left to future work. In this sense, we are providing the kinematical
background for the construction of a dynamical model. 
 
The main result of this paper will be to provide a consistent algebra
of generalised diffeomorphisms based on E$_9$ together with consistency checks using a generalised Scherk--Schwarz reduction. The
coordinates lie in the simplest $\fe_9$ highest weight
representation, sometimes called the ``basic'' or ``fundamental''
representation \cite{West:2003fc,Samtleben:2007an,Palmkvist:2013vya},
that can be identified with the Hilbert space of a
CFT on the E$_8$ lattice~\cite{Goddard:1986bp} and whose construction
will be reviewed in algebraic terms below. Due to the Hilbert space
structure, it will prove very convenient to employ Dirac notation to
write elements in this representation, its dual and tensor products.  

We will show in this paper that the full Lie derivative can be put in
a remarkably compact form 
\begin{align}
\mathcal{L}_{\xi,\Sigma}\, |{V}\rangle = 
 \langle{\partial}{}_V|{\xi}\rangle   |{V}\rangle 
 + \langle {\partial}{}_\xi | 
(C_0-1) |{\xi}\rangle  \otimes |{V}\rangle
 +\langle {\pi_\Sigma} | {C}_{-1} |{\Sigma}\rangle \otimes | {V}\rangle 
 \;,
 \label{fullLie0}
\end{align}
acting on a fundamental vector $|V\rangle$ with the rescaled coset
Virasoro generators  
$C_n\!\equiv\! 32\,L_n^{\rm coset}$~\cite{Goddard:1984vk}, acting on
tensor products of fundamental representations.  
The gauge parameters
combine a fundamental vector $| \xi \rangle$ as the generic
diffeomorphism parameter  
together with a two-index tensor which
we denote as $\Sigma \equiv | \Sigma \rangle \langle \pi_\Sigma |$ and
which is constrained 
in its second index as we specify below. 
The latter parameter is required for closure of the algebra, in analogy to a
similar term in the E$_8$ 
exceptional field theory with three external
dimensions~\cite{Hohm:2014fxa,Cederwall:2015ica}. This additional
gauge transformation in \eqref{fullLie0} does not absorb the standard diffeomorphism acting on the highest weights components of the vector field $|V\rangle$, and will therefore only gauge away unphysical components of the generalised vielbein in the exceptional field theory. Generalised diffeomorphisms based on the infinite-dimensional Kac--Moody algebra $\mf{e}_{11}$ have been proposed in~\cite{West:2014eza} up to an unknown connection.  The section constraint and the extra constrained gauge parameter $\Sigma$ that we crucially need for the closure do not feature in the proposal of~\cite{West:2014eza}, whereas we believe that they will be needed for the closure of the algebra. 

The transformations~\eqref{fullLie0} close into an algebra, provided
we impose the section constraint 
 \begin{align}
\langle \partial_1 | \otimes \langle \partial_2 |  \, ( C_0 - 1 +
\sigma )&=0 \ , \nonumber\\ 
 \langle \partial_1 | \otimes \langle \partial_2 |  \,  C_{-n} &=0\;,
 \quad \forall n>0\;,  \label{SectionConstraint0}\\ 
\left(  \langle \partial_1 | \otimes \langle \partial_2 |  +  \langle
\partial_2 | \otimes \langle \partial_1 |\right) \,C_1 &=0
\;.\nonumber 
 \end{align}
where $\sigma$ is the operator that exchanges the two factors of the
tensor product $\langle \partial_1|\otimes \langle \partial_2|$.
This is a special case of a general expression for the section
condition that applies in all extended geometries,
\bea
\bra{\partial_1}\otimes\bra{\partial_2}
\left[-\eta_{AB}T^A\otimes T^B+(\lambda,\lambda)+\sigma-1\right]=0\;.
\eea

After a generalised Scherk--Schwarz reduction with an appropriate Ansatz for the gauge parameters $|\xi\rangle$ and $\Sigma$ and the vector $|V\rangle$, the generalised diffeomorphisms
(\ref{fullLie0}) reduce to an algebraic action 
which precisely reproduces the gauge structure of
two-dimensional gauged supergravity~\cite{Samtleben:2007an}. The section constraints above 
then imply the quadratic constraints on the two-dimensional embedding tensor.

Remarkably, the entire construction appears to make little use of the explicit structure of E$_8$
and its specific tensor identities, in marked contrast to the analogous constructions for
the finite dimensional groups \cite{Berman:2012vc,Hohm:2014fxa}. Rather, most of the consistency of
the diffeomorphism algebra is a consequence of the underlying coset Virasoro symmetry. It is thus
natural to expect that the present construction is not limited to the case of E$_8$ and its affine extension
but naturally generalises to other affine algebras. We show that this is indeed the case.

Section \ref{E9Section} reviews some basic facts about $\fe_9$ and its
representations, including in particular some tensor products, and the
construction of coset Virasoro generators. In section
\ref{CoordinateSection}, we introduce coordinates and derivatives and
deduce the form of the section constraint using the coset Virasoro
generators.
Generalised diffeomorphisms, including ``extra'' local $\fe_9$-transformations, are introduced in section
\ref{DiffeoSection}, and are shown to close when acting on vectors.
Section \ref{sec:GSS} deals with the generalised Scherk--Schwarz
reduction, and shows that the diffeomorphisms reproduce the correct
structures, both for standard and non-Lagrangian gaugings, of
two-dimensional gauged supergravity. In section \ref{GeneralisationSection},
it is first shown how our results are generalised to other affine
algebras, and then how a completely general expression, valid for
arbitrary Kac--Moody algebras and highest weight coordinate
representations, for the generalised Lie derivative and the section
constraint can be derived. 
We conclude with a summary and discussion of our results and
indicate some questions for future research in section \ref{ConclusionsSection}.


\section{\texorpdfstring{E$_9$}{E9}: algebra and representations\label{E9Section}}


Here we review the structure of the affine algebra $\fe_9$ and some useful
facts about its representations. 
We denote by $\mf{e}_9$ the centrally extended loop algebra over
$\mf{e}_8$, together with the derivation generator~$\dL$. The
generators are 
\begin{align}
\mf{e}_9 = \left\langle T^A_m \,:\, A=1,\ldots, 248,\quad
m\in\mathds{Z} \right\rangle \oplus \mathds{R} \dK \oplus \mathds{R}
\dL  \,. 
\label{e9}
\end{align}
The first part is the loop algebra, $\dK$ is the central element and
$\dL$ the derivation acting by  
$\lb \dL, T^A_m\rb =-m T^A_m$. The remaining commutators are
\bea{}
\left[T^A_m,T^B_n\right]=f^{AB}{}_C\,T^C_{m+n}+\eta^{AB}m\,\delta_{m+n,0}\,\dK\komma
\label{ENineAlgebra}
\eea
with $\fe_8$ structure constants $f^{AB}{}_C$ and Killing metric
$\eta^{AB}$, and where 
the standard normalisation is used, so that
$f^{AC}{}_Df^{BD}{}_C=2\dcox\eta^{AB}=60\,\eta^{AB}$. The horizontal
$\mf{e}_8$ subalgebra of $\mf{e}_9$ is generated by the $T^A_0$ as
usual. 

\begin{figure}[hbt]
   \centering
   \includegraphics[width=8cm]{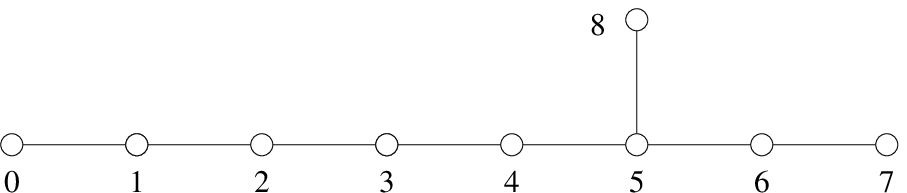}
   \caption{{\small The Dynkin diagram of $\fe_9$.}}
   \label{DynkinFigure}
\end{figure}
 
The algebra $\mf{e}_9$ admits highest and lowest weight
representations. Highest weight representations $R(\Lambda)$ are labelled by a
dominant integral weight $\sum_{i=0}^8 \ell^i \Lambda_i$ where the
labels are those of figure~\ref{DynkinFigure} and are distinguished by
their ``level'' $k$ which is the eigenvalue of the generator $\dK$
acting on the module. The level of $R(\Lambda)$ is
$k=\sum_{i=0}^8a_i\ell^i$, where $a_i$ are the Coxeter labels
$(a_0,\ldots,a_8)=(1,2,3,4,5,6,4,2,3)$. The leading states of $R(\Lambda)$
form the $\fe_8$ representation $r(\lambda)$ with highest weight 
$\lambda=\sum_{i=1}^8\ell^i\lambda_i$. Any dominant
integral highest weight can be shifted by an arbitrary real amount $-h
\delta$, where $\delta$ is the lowest positive null root of the affine
algebra and 
dual to the derivation $\dL$. This means that the $\dL$ eigenvalue on a
weight $\Lambda= \sum_{i=0}^8 \ell^i \Lambda_i-
h \delta$ is $h$.
The lowest weight module conjugate to the highest weight module $R(\Lambda)$ will be denoted by
$\overline{R(\Lambda)}$. 

At $k=1$ there is (up to $\delta$ shifts) only one dominant weight
$\Lambda_0=(100000000)$ and the corresponding module is called the
``basic'' representation of $\mf{e}_9$. By extrapolation of the
coordinate representations of E$_n$ exceptional geometries
(see \eg\ \cite{Berman:2012vc}),
one would expect this to be the right representation for the E$_9$
coordinates and we will show from different angles that this is indeed
the case.
When one constructs the
relevant invariant tensors used in the generalised
diffeomorphisms and appearing in the section condition, it is
important to have control over tensor products of highest weight
states, especially $R(\Lambda_0)$'s.\footnote{Highest weight modules of affine Kac--Moody algebras are closed under the tensor product operation, since they belong to ``category $\mathcal{O}$''~\cite{Kac:1990gs}, and tensor products are completely reducible but infinitely so, see also~\cite{Chari:1987}.}
Using the affine grading of (\ref{ENineAlgebra}), the module $R(\Lambda_0)$
is generated by 
acting with the generators on an $\mf{e}_8$ invariant scalar highest
weight state $\ket0$ 
satisfying 
\bea
T^A_n\ket0=0\komma\quad n\geq0\,,\quad \dL \ket0 =0\,,\quad (\dK-1)\ket0
=0\punkt 
\label{ROneZero}
\eea
The basic null states in the module appear as 
\bea
\mathbb{P}_{\bf{(27\,000)}}{}^{AB}{}_{CD}\,T^C_{-1}T^D_{-1}\ket0
\;,\label{null}
\eea 
which is straightforward to verify using (\ref{ENineAlgebra}) and
(\ref{ROneZero}) 
together with the projection operators on the tensor product of two
$\fe_8$ adjoint representations given in appendix~\ref{app:e8reps},
where also the 
dimensionalities of some $\mf{e}_8$ representations are listed.

The first few levels of $R(\Lambda_0)$ are 
\begin{align}
R(\Lambda_0)&={\bf1}_0\oplus{\bf248}_{-1}\oplus({\bf1}\oplus{\bf248}
\oplus{\bf3\,875})_{-2}
\oplus({\bf1}\oplus2\cdot{\bf248}\oplus{\bf3\,875}\oplus{\bf30\,380})_{-3}
\nonumber\\ 
&\quad{}
\oplus(2\cdot{\bf1}\oplus3\cdot{\bf248}\oplus2\cdot{\bf3\,875}\oplus{\bf30\,380}
        \oplus{\bf27\,000}\oplus{\bf147\,250})_{-4}\nonumber\\
&\quad{}
\oplus(2\cdot{\bf1}\oplus5\cdot{\bf248}\oplus3\cdot{\bf3\,875}
\oplus3\cdot{\bf30\,380}
      \oplus{\bf27\,000}\oplus{\bf147\,250}\oplus{\bf779\,247})_{-5}\nonumber\\ 
&\quad{}
\oplus\ldots\;.
\label{RL0}
\end{align}
The subscripts in the above equation refer to minus the number of
times the lowering generator of node $0$ where used. Equivalently, it
is minus the eigenvalue of the operator $\dL$, and we refer to the
subscript as ``affine level''.
We shall sometimes denote isomorphic
modules with shifted  
affine level
$h$ for the vacuum by
$R(\Lambda_0)_{-h}$. They satisfy $\dL \ket0 = h \ket0$. 
The character for $R(\Lambda_0)$ that counts only affine level
(where a term $c_nq^n$ corresponds to $c_n$ states at level $-n$)
has a
remarkable
form \cite{KacAddendum,LepowskyJ} in terms of the modular invariant
function $j$: 
\bea
\chi_{R(\Lambda_0)}(q)=\left(q\,j(q)\right)^{1/3}\;,
\eea
and we discuss this Hilbert space in some more detail in
appendix~\ref{FormalTrace}. 

In a grading with respect to the exceptional root (the simple root corresponding to node 8), the
fundamental representation 
has the following expansion in terms of $\fsl(9)$ representations,
\bea
R(\Lambda_0)&=&(10000000)_0\oplus(00000010)_{-1}\oplus(00010000)_{-2}\nonumber\\ 
&&{}\oplus\left[(10000000)\oplus(01000001)\right]_{-3}\nonumber\\ 
&&{}\oplus\left[(00000010)\oplus(00000002)\oplus(10000100)\right]_{-4}
\nonumber\\ 
&&{}\oplus\left[(00010000)\oplus(10100000)\oplus(00001001)\right]_{-5}
\nonumber\\ 
&&{}\oplus\left[2(10000000)\oplus2(01000001)\oplus(20000001)
     \oplus(00100010)\right]_{-6}\nonumber\\ 
&&{}\oplus\ldots\,,
\label{SLNineGrading}
\eea
while the adjoint is
\bea
\adj=\bigoplus\limits_{n\in\ZZ}
\left[(00100000)_{3n+1}\oplus(10000001)_{3n}\oplus(00000100)_{3n-1}\right]\oplus
2(00000000)_0 
\punkt
\eea
In these equations, the subscript is now given by minus the number of
times the lowering generator of node $8$ was used. 
Such a grading is suitable for analysing explicit solutions of the
section condition.  
We will however mostly use the affine grading, mainly because it is
better adapted to the Virasoro generators.

When dealing with representations of affine algebras, it is convenient
to use their CFT or current algebra interpretation. The Sugawara
construction \cite{Sugawara:1967rw} implies the 
presence of a Virasoro algebra, with generators
\bea
L^{(k)}_m=\frac{1}{2(k+\dcox)}\sum\limits_{n\in\ZZ}\eta_{AB}:T^A_nT^B_{m-n}:
\komma
\label{Sugawara}
\eea
and central charge $c_k=\frac{k\dim\fg}{k+\dcox}$. The dual Coxeter
number $\dcox$ for $\mf{e}_8$ is $\dcox=30$ and the colons refer to
standard normal ordering moving positive mode numbers to the
right. The Sugawara--Virasoro generators satisfy the commutation
relations 
\bea{}
\left[L^{(k)}_m,L^{(k)}_n\right]=(m-n)L^{(k)}_{m+n}+\frac{c_k}{12}
(m^3-m)\,\delta_{m+n,0} \,
\;,
\label{cl1}
\eea
and
\bea{}
\left[L^{(k)}_m,T^A_n\right]=-n\,T^A_{m+n}\komma
\label{cl2}
\eea
with the loop algebra. Often, we will not write the level $k$
superscript when the module is clear from the context in order to keep
the notation light.

In an irreducible highest weight representation
one can relate $L_0$ to the derivation operator $\dL$.
The eigenvalue of $L_0$ is given by the Sugawara construction
(\ref{Sugawara}), where $L_0$ reduces to $\frac{1}{k+\dcox}$ times the $\fe_8$ quadratic Casimir, whereas the eigenvalue of $\dL$
on the highest weight state can be shifted regardless of the
weight of the centrally extended loop algebra. (However, in the non-highest weight representation on the centrally
extended loop algebra itself, $L_0$ and $\dL$ act in the same way, and can be identified which each other in the further
extension to the affine algebra).

At $k=1$, we have $c_1=8$, and the highest weight state has
$h=0$, where $h$ is the $L_0$ eigenvalue.
At $k=2$, there are three irreducible highest weight representations,
namely
$R(2\Lambda_0)$, $R(\Lambda_7)$ and $R(\Lambda_1)$.
The leading $\mf{e}_8$ levels are
\bea
R(2\Lambda_0)&=&{\bf 1}_0\oplus{\bf 248}_{-1}\oplus({\bf 1}\oplus{\bf
  248}\oplus{\bf 3\,875} 
\oplus{\bf 27\,000})_{-2}\nonumber\\
&&{}\oplus({\bf 1}\oplus3\cdot{\bf 248}\oplus{\bf 3\,875}\oplus{\bf 27\,000}
\oplus2\cdot{\bf 30\,380}
\oplus{\bf 779\,247})_{-3}\oplus\ldots\;,\nonumber\\[1ex]
R(\Lambda_7)&=&{\bf 3\,875}_0\oplus({\bf 248}\oplus{\bf 3\,875}
\oplus{\bf 30\,380}\oplus{\bf 147\,250})_{-1}
\oplus\ldots\;,\nonumber\\[1ex]
R(\Lambda_1)&=&{\bf 248}_0\oplus({\bf 1}\oplus{\bf 248}\oplus{\bf 3\,875}
\oplus{\bf 30\,380})_{-1}
\nonumber\\
&&{}\oplus({\bf 1}\oplus3\cdot{\bf 248}\oplus2\cdot{\bf
  3\,875}\oplus{\bf 27\,000} 
\oplus2\cdot{\bf 30\,380}\oplus{\bf 147\,250}\oplus{\bf 779\,247})_{-2}
\nonumber\\
&&{}\oplus(2\cdot{\bf 1}\oplus6\cdot{\bf 248}\oplus5\cdot{\bf 3\,875}
\oplus3\cdot{\bf 27\,000}
\oplus5\cdot{\bf 30\,380}\oplus3\cdot{\bf 147\,250}\nonumber\\
&&{}\qquad\oplus3\cdot{\bf 779\,247}
\oplus{\bf 2\,450\,240}\oplus{\bf 4\,096\,000}\oplus{\bf 6\,696\,000})_{-3}
\oplus\ldots
\;.
\eea
The value of the Virasoro central charge at $k=2$ is $c_2=\frac{31}{2}$,
so tensor products of two $R(\Lambda_0)$'s must also
contain ``compensating'' Virasoro modules with $c=2c_1-c_2=\frac12$. 
This is within the minimal series~\cite{CFTbook} (with $m=3$, the Ising model), where 
\bea
c&=&1-\frac{6}{m(m+1)}\komma\quad m=3,4,\ldots\;,\nonumber\\
h^m_{r,s}&=&\frac{((m+1)\,r-ms)^2-1}{4m(m+1)}\komma
        \quad r=1,\ldots m-1\komma\quad s=1,\ldots r\;.
\eea
We can easily read off the eigenvalue $h$ of $L_0$ on the highest weight
states of the three representations, since
the values of the $\fe_8$ quadratic Casimir $C_2(r(\lambda))$, 
normalised to $\dcox$ in the adjoint representation, can be
calculated as  $C_2(r(\lambda))=\frac12(\lambda,\lambda+2\varrho)$.
They are
$0$, $48$, and $30$ in the three representations $r(0)={\bf 1}$, $r(\lambda_7)={\bf 3\,875}$
and $r(\lambda_1)={\bf 248}$, leading to $h=0$,
$\frac32$, and $\frac{15}{16}$, respectively.  
These values must be matched 
(see {\it e.g.} \cite{KacMoodyWakimoto}) by the possible values of $h^3_{r,s}$,
which are
$h^3_{1,1}=0$, $h^3_{2,1}=\frac12$, $h^3_{2,2}=\frac1{16}$.
There is the possibility of shifting with an integer, since the
eigenvalue of $\dL$ on a highest weight state can be
shifted. Conservation of $h$ leads to possible matchings
$0=0+h^3_{1,1}$, $2=\frac32+h^3_{2,1}$, $1=\frac{15}{16}+h^3_{1,1}$. This
shows that the first appearances of $R(2\Lambda_0)$, $R(\Lambda_7)$ and
$R(\Lambda_1)$ in $R(\Lambda_0) \otimes R(\Lambda_0)$ may be (with some integer
multiplicity) at affine levels $0$, $-2$ and $-1$, respectively. It thus suffices to
check the tensor product to affine level $-2$ in order to establish the
(integer) coefficients, which all turn out to be 1, such that
\cite{KacWakimoto1986}  
\begin{align}
R(\Lambda_0)\otimes R(\Lambda_0)
=\Vir^3_{1,1}\otimes R(2\Lambda_0)_0
\oplus\Vir^3_{2,1}\otimes R(\Lambda_7)_{-3/2}  
\oplus\Vir^3_{2,2}\otimes R(\Lambda_1)_{-15/16}\komma
\label{TensorTwoROne}
\end{align}
where $\Vir^m_{r,s}$ are Virasoro modules, keeping track of the
repeated occurrence of the three representations in $R(\Lambda_0) \otimes R(\Lambda_0)$. 
It is also easily checked that the first two terms in
(\ref{TensorTwoROne}) represent the symmetric product and the last one
the antisymmetric product.

The corresponding Virasoro characters are
\bea
\chi^3_{1,1}&=&\frac12
\left(\frac{\phi(q)^2}{\phi(\sqrt q)\phi(q^2)}
+\frac{\phi(\sqrt q)}{\phi(q)}\right)\nonumber\\
&=&1\phantom{+q}+q^2+q^3+2q^4+2q^5+3q^6+3q^7+5q^8+5q^9+7q^{10}+8q^{11}\nonumber\\
&&{}+11q^{12}+12q^{13}+16q^{14}+18q^{15}+23q^{16}+O(q^{17})\komma\nonumber\\[1ex]
\chi^3_{2,1}&=&\frac12
\left(\frac{\phi(q)^2}{\phi(\sqrt q)\phi(q^2)}
-\frac{\phi(\sqrt q)}{\phi(q)}\right)\nonumber\\
&=&\sqrt
q\,\Big(1+q+q^2+q^3+2q^4+2q^5+3q^6+4q^7+5q^8+6q^9+8q^{10}+9q^{11}\nonumber\\ 
&&{}\qquad\quad+12q^{12}+14q^{13}+17q^{14}+20q^{15}+25q^{16}+O(q^{17})\Big)
\komma\nonumber\\[1ex]
\chi^3_{2,2}&=&\frac{q^{1/16}\,\phi(q^2)}{\phi(q)}\nonumber\\
&=&q^{1/16}\,\Big(1+q+q^2+2q^3+2q^4+3q^5+4q^6+5q^7+6q^8+8q^9+10q^{10}+12q^{11}
\nonumber\\
&&{}\qquad\quad+15q^{12}+18q^{13}+22q^{14}+27q^{15}+32q^{16}+O(q^{17})\Big)\komma
\label{VirCharExp}
\eea
where $\phi(q)=\prod_{n=1}^\infty(1-q^n)$.
Note the absence of states at level $-1$ in the first of these
representations, which of course derives from the ${\rm SL}(2)$-invariance
of the highest weight state. This property will become important later.
The characters satisfy 
\bea
\left((\chi^3_{1,1})^2-(\chi^3_{2,1})^2\right)\chi^3_{2,2}=q^{1/16}\punkt
\eea
The coset Virasoro generators acting on (\ref{VirCharExp}) are given by
\bea
L_n^{\rm coset}&\equiv& 
\id\otimes L_n^{(1)} + L_n^{(1)} \otimes\id - L_n^{(2)}
\;,
\label{coset}
\eea
in terms of the level 1 and level 2 Virasoro--Sugawara operators (\ref{Sugawara}),
as a particular case of the coset construction~\cite{Goddard:1984vk}.
We will in the following often make use of the following rescaled
coset Virasoro  
generators:
\bea
C_n ~\equiv~32\,L_n^{\rm coset} &=&
32\left(
\id\otimes L_n^{(1)} + L_n^{(1)} \otimes\id - L_n^{(2)}\right)
\label{32coset}\\
&=& 
\id\otimes L_n^{(1)} + L_n^{(1)} \otimes\id
-\sum\limits_{p\in\ZZ}\eta_{AB}T^A_p\otimes T^B_{n-p}
\;.
\nonumber
\eea
A general coset Virasoro generator, acting on a tensor product of
states at $k=k_1$ and $k=k_2$, is
\bea
\label{eq:cosetL}
L^{{\rm coset}}_n&=&L_n^{(k_1)}\otimes\id+\id\otimes L_n^{(k_2)}
-L^{(k_1+k_2)}_n\nn\\
&=&\frac{1}{k_1+k_2+\dcox}\Bigl(L_n^{(k_1)}\otimes\dK
+\dK\otimes L_n^{(k_2)}-\sum\limits_{p\in\ZZ}\eta_{AB}T^A_p\otimes T^B_{n-p}
\Bigr)\\
&\equiv&-\frac{1}{k_1+k_2+\dcox}\,
\eta_{(n){\cal A}{\cal B}}\,T^{{\cal A}} \otimes T^{{\cal B}}\,,\nn
\eea
where the indices ${\cal A}, {\cal B}$ in the last expression run over
the semi-direct 
sum of the centrally extended loop algebra
with the full Virasoro algebra 
(although in this case the expression is zero whenever $\cal A$
or $\cal B$ corresponds to a Virasoro generator different from $L_n$),
and the (non-invertible) 
bilinear forms $\eta_{(n){\cal A}{\cal B}}$ are defined by this equation
and invariant under the loop algebra of $\mf{e}_9$. For $n=0$ we get
the standard invariant 
form on $\mf{e}_9$ (if we identify $L_0$ with $\dL$).

By construction the operators (\ref{32coset}) satisfy the Virasoro
algebra up to a factor of $32$ and for central charge
$\frac{1}{2}$. What will be important in the following is the algebra
they satisfy when acting on different $R(\Lambda_0)\otimes
R(\Lambda_0)$ subspaces of the level $3$ tensor product
$R(\Lambda_0)\otimes R(\Lambda_0)\otimes R(\Lambda_0)$.  
Let us consider the action of  (\ref{32coset}) on two factors of this
triple tensor  
product, where we use the notation
\bea
\overset{12}{C}_n &\equiv& -\eta_{(n){\cal A}{\cal B}}\,T^{{\cal A}}
\otimes T^{{\cal B}} \otimes \id\;,\nonumber\\ 
\overset{13}{C}_n &\equiv& -\eta_{(n){\cal A}{\cal B}}\,T^{{\cal A}}
\otimes \id\otimes  T^{{\cal B}}\;,\qquad \mbox{etc.}\;. 
\eea
Straight-forward computation then shows the following structure
\begin{align}
\label{Commut} \lb \Kmat{13}_m , \Kmat{23}_n  \rb = \frac{m-n}{2}
\left(  \Kmat{13}_{m+n} +  \Kmat{23}_{m+n} - \Kmat{12}_{m+n} \right) +
\frac23 m(m^2-1)\, \delta_{m+n,0}  +  \Kmat{123}_{m+n}\,, 
\end{align}
where the last operator is defined as
\begin{align}
  \Kmat{123}_m &= \sum_{p,q \in\ZZ} f_{ABC}\, T^A_p \otimes T^B_q
  \otimes T_{m-p-q}^C \nn\\ 
 &\quad + \sum_{p\in\ZZ} \left(\frac{m}{2}-p\right) \eta_{AB} \left( \id\otimes
  T^A_p \otimes T_{m-p}^B + T_{m-p}^A \otimes \id \otimes T^B_{p} +
  T^A_p\otimes T^B_{m-p}\otimes \id \right)\;, 
\end{align}
and completely antisymmetric under the exchange of the three
spaces. It can be written in compact form as 
\bea
\Kmat{123}_{m} &=& f^{{\cal A}{\cal B}}{}_{\cal C}\, 
\eta_{(m){\cal A}[{\cal D}}\, \eta_{(0){\cal E}]{\cal B}}
\;{T}{}^{\cal D} \otimes {T}{}^{\cal E} \otimes {T}{}^{\cal C} 
\;
\eea
with the bilinear forms $\eta_{(n){\cal AB}}$ from (\ref{eq:cosetL})
and structure constants $f^{{\cal A}{\cal B}}{}_{\cal C}$ 
combining (\ref{ENineAlgebra}),(\ref{cl1}), (\ref{cl2}). 
Using its antisymmetry one can show the following relations between commutators
\begin{align} \label{FirstCom} \lb \Kmat{13}_m , \Kmat{23}_n  \rb  -
  \lb \Kmat{12}_m , \Kmat{13}_n  \rb  
&= (m-n) \,\big(  \Kmat{23}_{m+n}- \Kmat{12}_{m+n}\big) \ , \\
\label{SeconCom} \lb \Kmat{23}_m , \Kmat{12}_n +\Kmat{13}_n \rb  &=
(m-n)\,\Kmat{23}_{m+n} + \frac{4}{3} m(m^2-1) \delta_{m+n,0} \ , \\ 
\label{ThirdCom}\lb \Kmat{13}_n , \Kmat{23}_{m-n}  \rb  - \lb
\Kmat{13}_p , \Kmat{23}_{m-p}  \rb  &= (n-p)\, \big(  \Kmat{13}_{m}+
\Kmat{23}_{m}- \Kmat{12}_{m}\big)  
+ \frac{2}{3} \big( n (n^2-1) - p (p^2-1)\big) \,\delta_{m,0} \;,
\end{align}
which will be useful in the following.

\section{Coordinates and section constraint\label{CoordinateSection}}

By extrapolation from the systematics of the coordinate representation
for E$_n$ one expects that the internal coordinates of E$_9$
exceptional field theory should transform in the fundamental
representation $R(\Lambda_0)$ of $\mf{e}_9$~\cite{West:2003fc,Samtleben:2007an,Palmkvist:2013vya}. We proceed with this
assumption and denote the coordinates as $Y^M$. As for the
finite-dimensional groups, consistency of the theory should require a
section  
constraint that eliminates the dependence of fields on all but the
physical coordinates. 
Derivatives $\partial_M$ transform in the dual
$\overline{R(\Lambda_0)}$ of the fundamental representation that
decomposes in analogy with~\eqref{RL0} according to 
\begin{align}
\label{eq:bas}
\overline{R(\Lambda_0)} = {\bf 1}_0 \oplus {\bf 248}_{1} \oplus
\left({\bf 1}\oplus {\bf 248}\oplus {\bf 3\,875}\right)_{2}\oplus\ldots 
\end{align}
under $\mf{e}_8\subset \mf{e}_9$. 

The section constraint is expected to be bilinear in derivatives, {\it
  i.e.}, to lie in the 
tensor product of $\overline{R(\Lambda_0)}\otimes
\overline{R(\Lambda_0)}$ which can be decomposed in analogy to
(\ref{TensorTwoROne}). 
The possible projectors onto $\mf{e}_9$ representations within this
tensor product are naturally expressed in terms 
of the coset Virasoro generators defined in (\ref{coset}). Our Ansatz
for the (strong) $\mf{e}_9$ section constraint is
\begin{align}
\langle \partial_1 |\otimes  \langle \partial_2 | \,(C_0-1+ \sigma ) =0
\label{sectionA}
\;.
\end{align}
Here and in the following we use a notation in which the fundamental
representation and its dual are represented 
by ket- and bra-vectors, respectively. 
In particular, derivatives $\partial_M$ are seen as bra-states in
lowest-weight modules at $k=-1$. 
Subscripts $_{1,2}$ on the derivatives indicate that these derivatives
may act on different objects. 
The operator $C_0$ is the rescaled coset Virasoro generator from
(\ref{32coset}), 
and $\sigma$ denotes the permutation operator on a tensor product
\bea
\langle \partial_1 |\otimes  \langle \partial_2 | \,\sigma &\equiv&
\langle \partial_2 |\otimes  \langle \partial_1 | 
\;.
\label{sigma}
\eea

We will show below that the Ansatz (\ref{sectionA}) is compatible with the expected
solutions of the section constraint. 
As a first check, let us verify that (\ref{sectionA}) indeed
reproduces the section constraints from three-dimensional  
E$_8$ exceptional field theory upon proper embedding.
Comparing the coordinates
to E$_8$ exceptional field theory with three external dimensions, we
expect the lowest singlet ${\bf 1}_{0}$ in the level decomposition  
(\ref{RL0}) to correspond to the singlet in the $3\longrightarrow2+1$
decomposition of external dimensions while the adjoint ${\bf
  248}_{-1}$ 
on the first level should correspond to the internal coordinates of
E$_8$ exceptional field theory. 
Restricting coordinates to these two lowest levels, {\it i.e.}, assuming
\bea 
\langle \partial | = \langle 0 | ( \partial_0 + T_1^A \partial_A ) \ , 
\label{der3D}
\eea
we can then evaluate the constraint (\ref{sectionA}) as
\bea 
0 &=&
 \langle \partial_1 |\otimes  \langle \partial_2 | (C_0-1+ \sigma )
 \nn \\[1ex] 
&=&  \langle 0 |\otimes  \langle 0 |\, \partial_{1A} \partial_{2B}\,  \Bigl( \,
\Pi^{AB}{}_{CD} \, T_1^C\otimes T_1^D - T_1^A T_1^B \otimes \id - \id
\otimes T_1^B T_1^A \Bigr ) \ ,  
\eea
where 
\bea 
\Pi^{AB}{}_{CD} \equiv 2\, \delta^{(A}_{C} \delta^{B)}_D - f^A{}_{CE}
f^{EB}{}_D = 14 \,(\mathbb{P}_{\bf 3\,875})^{AB}{}_{CD}  
+ 4\, \eta^{AB} \eta_{CD} - 2\, f^{AB}{}_E f^E{}_{CD} \; 
\label{Pi}
\eea
is given as a linear combination of projectors onto the ${\bf 1}$,
${\bf 248}$ and ${\bf 3\,875}$, cf.~(\ref{EEightProjectionOperators}). 
Using the property that $\langle 0 | \, T_1^A T_1^B$ is only non-zero
for $(AB)$ in the ${\bf 1} \oplus {\bf 248} \oplus {\bf 3\,875}$,  
cf.~(\ref{null}), one recovers the E$_8$ section constraint
\cite{Hohm:2014fxa} 
\bea
\partial_A \otimes \partial_B\left(
\mathbb{P}_{\bf 1}+\mathbb{P}_{\bf 248}+\mathbb{P}_{\bf 3\,875}
\right)^{AB}{}_{CD}  =0
\;.
\label{section3D}
\eea
In turn, one observes that with derivatives $\partial_A$ constrained
by (\ref{section3D}),  
the tensor product of two derivatives (\ref{der3D}) is exclusively
contained in the  
the leading $\overline{R(2\Lambda_0)}_0$ and the leading
$\overline{R(\Lambda_1)}_1$ 
in the expansion (dual to) (\ref{TensorTwoROne}). 
The full $\fe_9$ section condition is then expected to be equivalent to the 
vanishing of the remaining (infinite number of) irreducible
representations in $\overline{R(\Lambda_0)} \otimes\overline{R(\Lambda_0)}$, among them all 
$\overline{R(\Lambda_7)}$'s.
As a simple consequence of the grading, all $L^{\rm coset}_m$, $m<0$
vanish when acting 
on products of (\ref{der3D}), so they may be included in the
(conjugate) section  
condition ``for free''. 
Moreover, the absence of level $-1$ states in 
Vir$^3_{1,1}$, cf.~(\ref{VirCharExp}), then implies that also $C_1$ 
annihilates these products.
Together, we arrive at the following proposal for the $\mf{e}_9$
section constraints 
\begin{subequations}
\label{SCall}
\begin{align} \label{SectionConstraint}\langle \partial_1 | \otimes
  \langle \partial_2 |  \, 
 ( C_0 - 1 + \sigma )&=0 \ , \\
 \label{SectionConstraintN} \langle \partial_1 | \otimes \langle
 \partial_2 |   \, C_{-n} &=0 \;,\quad \forall n>0\;, \\ 
 \label{SectionConstraintP}\left(  \langle \partial_1 | \otimes
 \langle \partial_2 |  +  \langle \partial_2 | \otimes \langle
 \partial_1 |\right) \,C_1 &=0 \;, 
 \end{align}
 \end{subequations}
which correctly reproduces the $D=3$, E$_8$ section constraint.
Moreover, we show in Section \ref{ArbitraryKMSection} that
\eqref{der3D} satisfying \eqref{section3D} is the unique solution to
\eqref{SCall} up to conjugation in E$_9$. 

There can be different definitions of E$_9$, in particular for the
space of functions defining the loop group. The proof of section
\ref{ArbitraryKMSection} uses the definition of a Kac--Moody group of
\cite{Peterson:1983} that corresponds in the affine case to taking the
loop group of meromorphic functions in $E_{8}$ with poles at zero and
infinity only. It follows by iterations that the maximal vector
spaces  in $R(\Lambda)$ of solutions to \eqref{SCall} are E$_9$
conjugate to the expected type IIB and eleven-dimensional supergravity
solutions. The latter can be seen explicitly in the $\fsl(9)$ level
decomposition (\ref{SLNineGrading}) of the coordinate representation,
for which a  
solution to the section constraints (\ref{SCall})  
is given by restricting the coordinate dependence to the $\fsl(9)$
vector on the lowest level, which corresponds to the nine coordinates that
allow to embed the full eleven-dimensional supergravity in exceptional field theory.

Although the constraints in
  (\ref{SCall}) are independent as algebraic equations, already
  the symmetric part of \eqref{SectionConstraint} is sufficient to imply that they are all satisfied. There is no clear consensus in the literature about what is
  to be called a section constraint (except that it should be strong
  enough). Sometimes, the complement to $\overline{R(2\Lambda_0)}$ in
  the symmetric product
  $\overline{R(\Lambda_0)}\otimes_s\overline{R(\Lambda_0)}$ is taken
  as the constraint. This is suitable in the
  context of {\it e.g.} the tensor hierarchy algebra
  \cite{Palmkvist:2013vya,Greitz:2013pua,Bossard:2017wxl,CarboneCederwallPalmkvist}.
Here, we choose to include all representations that vanish in the
section, also antisymmetric ones.

In addition to reproducing the expected physical solutions, the main
and defining 
characteristics of the proper section constraints is the fact that
they should guarantee closure 
of the algebra of generalised diffeomorphisms. This is what we will
show in the next section. 

\section{Generalised diffeomorphisms\label{DiffeoSection}}

Having identified a reasonable set of section constraints
(\ref{SCall}), we will now establish the algebra of generalised
diffeomorphisms. 
For the finite-dimensional groups, the generic action of a generalised
diffeomorphism 
on a vector field is of the form~\cite{Coimbra:2011ky,Berman:2012vc}
\bea
\LL_\xi V^M=\xi^N\partial_NV^M+Z^{MN}{}_{PQ}\,\partial_N\xi^PV^Q
\; , 
\label{GenLieDer0}
\eea
with an invariant tensor $Z^{MN}{}_{PQ}$ which up to a possible weight
term is built 
from the projector onto the adjoint representation
\bea
Z^{MN}{}_{PQ} &=& -\alpha\,\mathbb{P}^M{}_Q{}^N{}_P +
\beta\,\delta_P{}^N\delta_Q{}^M 
\;,
\eea
and is unique up to two constants $\alpha$ and $\beta$\,. With a vector field we mean a vector that could be a gauge transformation parameter $\xi$; we do not consider vectors of different weight.
For $\fe_9$,
the natural candidate 
for  this tensor is thus given by
\bea
Z^{MN}{}_{PQ}&=&\alpha\left(
\sum\limits_{n\in\ZZ}\eta_{AB}(T^A_n)^M{}_Q(T^B_{-n})^N{}_P
-\delta^M{}_Q\,(L_0)^N{}_P-(L_0)^M{}_Q\,\delta^N{}_P
\right)\nonumber\\
&&{}+\beta\,\delta^M{}_Q\,\delta^N{}_P\punkt
\label{Zcand}
\eea
It is important that $Z^{MN}{}_{PQ}$ (up to a possible scaling) 
is $\fe_9$ valued in the pairs ${}^M{}_Q$ and
${}^N{}_P$. In the following we will often turn to an index-free notation
in which (\ref{Zcand}) takes the compact form
\bea
Z =\sigma\,\left( -\alpha\,C_0 +\beta\right)
\;,
\label{Zcomp}
\eea
with the permutation operator $\sigma$ from (\ref{sigma}) and the
rescaled coset Virasoro generator $C_0$ from (\ref{32coset})\,.
The coefficients $\alpha$, $\beta$ are usually determined from closure
of the algebra of transformations (\ref{GenLieDer0}), for which a crucial
role is played by the fact that the section constraint of the theory
ensures the vanishing of~\cite{Berman:2012vc}
\bea
\langle \partial_1 |\otimes  \langle \partial_2 | \, Y =0
\;,
\label{sectionY}
\eea
for the tensor $Y\equiv Z+1$\,,
 \ie, $Y$ has to be a linear
combination of projections on irreducible representations in the
section condition. In the present case this will be an infinite number
of representations.
Comparing (\ref{Zcomp}) and (\ref{sectionY}) to the section constraints
(\ref{SCall}) identified in the
previous section, we read off the values $\alpha=\beta=-1$, for which
\bea
Y=\sigma(C_0+\sigma-1)\;.\label{E9YTensor}
\eea
In particular, this implies that the canonical weight of a vector is
$\beta=-1$.  With `canonical weight' (sometimes also called `distinguished weight' in the literature) we mean the weight of the gauge parameter $\xi$.
For E$_d$ exceptional field theory it is $\beta =
\frac{1}{9-d}$, which would diverge for $d=9$, but we shall see in section \ref{ArbitraryKMSection} that the appropriate definition for the highest weight coordinate module $R(\lambda)$ is $\beta = (\lambda,\lambda)-1$ that gives indeed $\beta=-1$ for E$_9$. A canonical co-vector (like {\it e.g.} a derivative) then has weight $\beta=+1$.

The tensors $Z$ and $Y$ (and thus the section constraint) can also be derived from extensions of $\fe_9$ in the same was as
in \cite{Palmkvist:2015dea} for finite-dimensional $\fe_d$. These extensions are the Lie algebra $\fe_{10}$
and a Lie superalgebra of Borcherds type, giving the antisymmetric and symmetric parts of $Y$, respectively. In both cases the algebra
is obtained by adding a node to the Dynkin diagram of $\fe_9$ (``white'' or ``gray''), and $\dL$ can be identified with the Cartan generator
corresponding to this additional node.

In the index-free notation, the generalised diffeomorphism
(\ref{GenLieDer0}) now reads 
\begin{align}
\LL_\xi\ket V =
 \langle{\partial}{}_V|{\xi}\rangle |{V}\rangle 
 + \langle {\partial}{}_\xi | 
(C_0-1) |{\xi}\rangle \otimes |{V}\rangle\,,
\label{OperatorDiff}
\end{align}
where the subscript on the derivatives indicate what they act on, {\it e.g.}
\bea  \langle{\partial}{}_V| \otimes  |{V}\rangle \otimes | \xi\rangle  = \biggl(  \Big\langle \frac{ \partial}{\partial Y}  \Big| \otimes  |{V(Y)}\rangle \biggr)  \otimes | \xi(Y)\rangle\ .  \eea
Specifically, our index-free conventions are such that for a tensor product 
one understands the bra and the ket states to be ordered from left to right, 
such that for example 
\begin{align}  
| \overset{2}W \rangle = \langle \overset{1}\omega| \overset{12}{X} |
\overset{1}{\xi} \rangle\otimes | \overset{2}{V} \rangle  
\qquad
\Longleftrightarrow\qquad
| W \rangle = \langle \omega| {X} | {\xi} \rangle\otimes |{V} \rangle\ ,  
\end{align}
corresponding to the following expression in indices
\begin{align}
W^M = \omega_N X^N{}_P{}^M{}_Q \,\xi^P V^Q\,.
\end{align}
Similarly, the labels on the states will be avoided in expressions of the type 
\begin{align}  
| \overset{3}W \rangle = \langle \overset{1}\omega| \otimes  \langle
\overset{2}\upsilon | \,\overset{12}{X}\,  \overset{23}{Y}\,  |
\overset{1}{\xi} \rangle\otimes | \overset{2}{\eta} \rangle \otimes |
\overset{3}{V} \rangle 
\quad
\Longleftrightarrow\quad
| W \rangle = \langle \omega| \otimes  \langle \upsilon |
\,\overset{12}{X}\,  \overset{23}{Y} \, | {\xi} \rangle\otimes |
          {\eta} \rangle \otimes | {V} \rangle\;, 
\end{align}
corresponding to the following expression in indices
\begin{align}  
W^M =  \omega_{N} \upsilon_P X^{N}{}_Q{}^P{}_R Y^R{}_S{}^M{}_T \xi^Q \eta^S V^T \ .
\end{align}

Having set up the notation, let us come back to the generalised
diffeomorphisms (\ref{OperatorDiff}).  
It comes as no surprise that the transformations (\ref{OperatorDiff})
do not close into an algebra. 
This is the case already for the generalised diffeomorphisms
associated with the algebra $\fe_8$ 
and it can be seen as a manifestation of the fact that
in three dimensions dual gravity degrees of freedom become part of the scalar 
sector~\cite{Coimbra:2011ky,Berman:2012vc}.
Yet, in this case a consistent symmetry algebra can be defined upon enlarging 
(\ref{GenLieDer0}) by local algebra-valued rotations with constrained gauge 
parameters~\cite{Hohm:2013jma,Hohm:2014fxa}.
The generic pattern in exceptional field theories for $\fe_d$ ({\it i.e.}, with $11-d$ external dimensions) 
is the appearance of additional covariantly constrained $(9-d)$-forms
in the dual fundamental representation. For E$_8$ exceptional field
theory these are the gauge fields whose associated 
gauge transformations are required for closure of the diffeomorphism algebra.
For E$_9$ exceptional field theory in contrast, one expects additional
fields to appear among the 
scalar fields, {\it i.e.} its scalar sector should carry not only a group
valued matrix ${\cal M}_{MN}$ but 
also 0-forms of type $\chi_M$ algebraically constrained by the 
section constraints (\ref{SCall}).
In the gauge sector we then expect vector fields $A_\mu{}^M$ in the
fundamental representation 
together with two-index gauge fields $B_{\mu}{}^N{}_M$ algebraically
constrained in  
its last index according to the section constraints. Their associated
gauge transformations with parameter 
$\Sigma^N{}_M$ are then responsible for closure of the full
diffeomorphism algebra.  
Fields of the same two-index structure appear
in E$_8$ exceptional field theory among the two-forms and are required
in order to close the algebra of gauge transformations and
supersymmetry on the vector
fields~\cite{Hohm:2014fxa,Baguet:2016jph}. 

In index-free notation, we will denote the new gauge parameter as
\bea
\Sigma^N{}_M &:&  |\Sigma\rangle \langle \pi_\Sigma |
\;,
\eea
to keep track of its two-index nature (keeping in mind that in general
this matrix is not factorised). 
The constrained nature of its first index is then expressed via (\ref{SCall}) as
\begin{align}  \langle \partial | \otimes \langle \pi_\Sigma |  \,
 ( C_0 - 1 + \sigma )&=0 \ , \nonumber\\
 \langle \partial | \otimes \langle \pi_\Sigma |   \, C_{-n} &=0
 \;,\quad \forall n>0\;, \nonumber\\ 
\left(  \langle \partial | \otimes \langle \pi_\Sigma |  +  \langle
\pi_\Sigma | \otimes \langle \partial |\right) \,C_1 &=0 \;.
                                 \label{SectionConstraintSigma}
 \end{align}
Combining (\ref{OperatorDiff}) with the new gauge transformations, we
arrive at the following 
definition for a generalised diffeomorphism, 
\begin{align}
\mathcal{L}_{\xi,\Sigma} |{V}\rangle = 
 \langle{\partial}{}_V|{\xi}\rangle |{V}\rangle 
 + \langle {\partial}{}_\xi | 
(C_0-1) |{\xi}\rangle \otimes |{V}\rangle
 +\langle {\pi_\Sigma} | {C}_{-1} |{\Sigma}\rangle \otimes | {V}\rangle 
 \;,
 \label{fullLie}
\end{align} 
with gauge parameters given by a vector field $|\xi\rangle$ and a
constrained tensor  
$|\Sigma\rangle \langle \pi_\Sigma |$ constrained by
(\ref{SectionConstraintSigma}).
The last term in (\ref{fullLie}) carries the coset Virasoro
generator $C_{-1}$ from 
(\ref{32coset}), such that it maps the $R(\Lambda_0)$ module to the
isomorphic module  
with an  $L_0$ spectrum  shifted by $1$, so that 
\begin{align} \label{densityL0} \dL   |\Sigma \rangle \langle
  \pi_\Sigma | = (L_0 +1)  |\Sigma \rangle \langle \pi_\Sigma |\ ,
  \qquad |\Sigma \rangle \langle \pi_\Sigma | \dL   =  |\Sigma \rangle
  \langle \pi_\Sigma |L_0 \ , \end{align}  
with $L_0$ being the Sugawara--Virasoro operator \eqref{Sugawara}. The weight of the gauge parameter $\ket \Sigma$ is $0$ in contrast to $\ket\xi$ that has weight $1$, such that in overall $\ket \Sigma \bra{\pi_\Sigma}$ has weight $-1$. 
More generally, one computes that the operator $C_n$ 
acting on the product of two vectors $|V\rangle$ and $|W\rangle$ of
canonical weight $-1$ 
shifts the weight from $-2$ to $n-2$:
\begin{align}
 &\phantom{=} C_n \,\cL_{\xi} \left(  |V\rangle \otimes |W\rangle \right)\nn\\
  & = \langle \partial_V+\partial_W | \xi\rangle  C_n |V\rangle \otimes
  |W\rangle+  \langle \partial_\xi | \,  \Kmat{23}_n\, ( \Kmat{12}_0+
  \Kmat{13}_0-2)\,  |\xi\rangle\otimes |V\rangle \otimes
  |W\rangle\nn\\  
 & = \langle \partial_V+\partial_W | \xi\rangle C_n  |V\rangle \otimes
  |W\rangle+  \langle \partial_\xi | \biggl(  \lb  \Kmat{23}_n  ,
  \Kmat{12}_0+ \Kmat{13}_0 \rb+ ( \Kmat{12}_0+
  \Kmat{13}_0-2)\, \Kmat{23}_n\, \biggr)   |\xi\rangle\otimes
  |V\rangle \otimes |W\rangle \nn\\  
  & = \langle \partial_V+\partial_W | \xi\rangle C_n |V\rangle \otimes
  |W\rangle+  \langle \partial_\xi | \, \bigl( \Kmat{12}_0+
  \Kmat{13}_0+n-2 \bigr)    |\xi\rangle\otimes( C_n
  |V\rangle \otimes |W\rangle )\ , 
\end{align}
where we have made use of (\ref{SeconCom}).
We recall that the weight appears in the generalised Lie derivative
(\ref{OperatorDiff}) as the integral shift of $C_0$. 

Note that in order to view the extra local rotations in the last term in (\ref{fullLie}) as an element ``in
the algebra'', the centrally extended loop algebra has to be supplemented by $L_{-1}$. This extension is (up to a sign convention)
the symmetry algebra $\mathfrak{G}$ used in \cite{Samtleben:2007an} to describe the structure of gauged supergravity in two dimensions,
which we will rederive from the generalised diffeomorphisms (\ref{fullLie}) in section~\ref{sec:GSS}.
Moreover, it agrees precisely with the level zero content of the
tensor hierarchy algebra corresponding to $\fe_9$, as defined in \cite{Bossard:2017wxl} for general $\fe_d$.
In general there is an additional highest weight module of generators, which reduces to the single element $L_{-1}$ for $d=9$. 

Before we address the closure of the algebra of transformations
(\ref{fullLie}), let us spell out the  
action on a co-vector of canonical weight 
\begin{align}
\mathcal{L}_{\xi,\Sigma} \langle \omega |  =  \langle \partial_\omega
| \xi\rangle \langle \omega |  
- \langle \partial_\xi| \otimes \langle \omega|\,(C_0-1)\, |\xi
\rangle
-\langle {\pi_\Sigma}|\otimes \langle \omega | {C}_{-1} |{\Sigma}\rangle
\;,
\end{align}
and note that if the co-vector $\langle \omega|$ is constrained by the
section constraint, such 
as the gauge parameter $\langle \pi_\Sigma|$ in
(\ref{SectionConstraintSigma}), it follows directly that 
\begin{align}
\mathcal{L}_{\xi,\Sigma} \langle \omega |  =  \langle \partial_\omega
| \xi\rangle \langle \omega |  
+ \langle \omega | \xi\rangle \langle \partial_\xi |
\ , 
\label{LieConstrained}
\end{align}
{\it i.e.}, also its resulting Lie derivative is constrained, and reduces to the ordinary Lie derivative.

Let us now check that the algebra of generalised diffeomorphisms
(\ref{fullLie}) closes. As a first step we compute 
the obstruction to the closure of the pure Lie derivative
$\mathcal{L}_{\xi,0} = \mathcal{L}_{\xi}$.   
We thus calculate
$([\LL_\xi,\LL_\eta]-\LL_{\leftbr\xi,\eta\rightbr})\ket V$, where
$\leftbr\xi,\eta\rightbr\equiv\frac12\,(\LL_\xi\eta-\LL_\eta\xi)$\,.
For $\fe_d$ with
$d\leq7$, this difference is $0$, and for $d=8$ it gives the ``extra'' local
$\fe_8$ transformation \cite{Hohm:2014fxa,Cederwall:2015ica}.
Let us go through the different types of terms arising. The terms with
two derivatives on $\ket V$ vanish trivially (due to antisymmetry under
$\xi\leftrightarrow\eta$). The terms with one derivative on $\ket V$ become
(here, antisymmetry between the parameters is implicit)
\bea
-\bra{\partial_\xi}\otimes\bra{\partial_V}\,
\big(\overset{12}{C}_0+\overset{12}{\sigma\vphantom{C}}
-1\big)\ket\xi\otimes\ket\eta\otimes\ket 
V\komma 
\eea
which vanishes thanks to the section condition (the superscripts on
$C_0$ and $\sigma$ 
indicate which pair of positions it acts on).

The terms without derivatives on $\ket V$ come in two groups, either the
two derivatives act on different parameters or on the same. When the two derivatives act on different gauge parameters, one obtains
\begin{align}
&\quad \frac12 \langle \partial_\xi | \otimes \langle \partial_\eta |
  \left( 2\left(\Kmat{13}_0-1\right)\left(\Kmat{23}_0-1\right)
  -\left(\Kmat{23}_0-1\right)\left(\Kmat{12}_0-1\right)
  -\overset{12}{\sigma}\left(\Kmat{23}_0-1\right)\right) |\xi\rangle
  \otimes |\eta\rangle\otimes|V\rangle\nn\\ 
&\hspace{10mm} - (\eta\leftrightarrow \xi)\nn\\
&=\frac12 \langle \partial_\xi | \otimes \langle \partial_\eta |
  \left[\Kmat{12}_0+\Kmat{13}_0, \Kmat{23}_0 \right] |\xi\rangle
  \otimes |\eta\rangle\otimes|V\rangle - (\eta\leftrightarrow
  \xi) =0\,,
\end{align}
where we have used the section constraint~\eqref{SectionConstraint} to
re-express $\overset{12}{\sigma}$ and used~\eqref{SeconCom}. The terms
with both derivatives on the same gauge parameter are the only
non-vanishing ones and can be arranged as 
\begin{align}
\Delta_{\xi,\eta} |V\rangle &\equiv \left( \left[\mathcal{L}_\xi ,
  \mathcal{L}_\eta\right] -\mathcal{L}_{\leftbr \xi,\eta\rightbr}\right) |
V\rangle\nn\\ 
&= \frac12\langle \partial_\eta | \otimes \langle \partial_\eta |
\left(- \Kmat{13}_0 + \Kmat{23}_0 - \Kmat{123}_0\right)  |
\xi\rangle\otimes | \eta\rangle \otimes |V \rangle \nn\\ 
& \qquad +\frac12 \langle \partial_\xi | \otimes \langle \partial_\xi
| \left(- \Kmat{13}_0 + \Kmat{23}_0 - \Kmat{123}_0\right)  |
\xi\rangle\otimes | \eta\rangle \otimes |V \rangle 
\;.
\end{align}
Now we shall show that this variation can be absorbed in 
a transformation of the type (\ref{fullLie}) with a constrained gauge
parameter $ |\Sigma \rangle \langle \pi_\Sigma |$.  
Using the identity \eqref{Commut} one shows that 
\begin{align} 
- \Kmat{13}_0 + \Kmat{23}_0 - \Kmat{123}_0 = \frac12  \lb
\Kmat{13}_{-1} - \Kmat{23}_{-1}  ,  \Kmat{12}_1 \rb \ .  
\end{align}
Substituting this into $\Delta_{\xi,\eta} |V\rangle$ one finds that
the term of the commutator with $ \Kmat{12}_1$ on the left vanishes
according to the section constraint (\ref{SectionConstraintP}), such
that the result can be written as  
\begin{align}
\Delta_{\xi,\eta} |V\rangle &=\frac14 \,\langle \partial_\eta | C_{-1}
\big( \langle \partial_\eta | C_1 | \eta \rangle \otimes | \xi \rangle
-  \langle \partial_\eta | C_1 | \xi \rangle \otimes | \eta
\rangle\big) \otimes | V \rangle \nn\\ 
& \qquad + \frac14\, \langle \partial_\xi | C_{-1} \big( \langle
\partial_\xi | C_1 | \eta \rangle \otimes | \xi \rangle -  \langle
\partial_\xi | C_1 | \xi \rangle \otimes | \eta \rangle\big) \otimes |
V \rangle  
\;.
\end{align}
We thus obtain closure of pure Lie derivatives into full generalised
diffeomorphisms (\ref{fullLie}) with the additional 
gauge parameter given by
\begin{align}
 | \Sigma \rangle \langle \pi_\Sigma| &\equiv 
 \frac14 \, \langle \partial_\eta | C_1 | \big(| \eta \rangle \otimes
 | \xi \rangle - |\xi \rangle \otimes | \eta \rangle\big)  \langle
 \partial_\eta |  
 + \frac14\,  \langle \partial_\xi | C_1 | \big( |\eta \rangle \otimes
 | \xi \rangle -| \xi \rangle \otimes | \eta \rangle\big)  \langle
 \partial_\xi | \ .  
 \end{align} 
 Note that this is manifestly constrained in its last index since the
 bra components are all partial derivatives.  

Next, we need to check that also the commutator of both kinds of
transformations in (\ref{fullLie}) closes into a gauge transformation  
\begin{align}
\label{eq:mixedcomm}
\lb \mathcal{L}_{\xi,0},\mathcal{L}_{0,\Sigma }\rb  |V\rangle &= 
\langle \partial_\Sigma + \partial_V |\xi\rangle \langle \pi_\Sigma |
C_{-1} |\Sigma\rangle \otimes |V\rangle \nn\\ 
&\quad + \langle \partial_\xi | \otimes \langle \pi_\Sigma
|\,(\Kmat{13}_0-1)\, \Kmat{23}_{-1}\,  
|\xi\rangle \otimes | \Sigma\rangle \otimes |V\rangle\nn\\
&\quad - \langle \partial_V | \xi\rangle \langle \pi_\Sigma | C_{-1} |
\Sigma\rangle \otimes |V\rangle - \langle\partial_\xi|\otimes
\langle\pi_\Sigma | \,\Kmat{23}_{-1}\, (\Kmat{13}_0-1)\, |\xi\rangle
\otimes |\Sigma\rangle\otimes |V\rangle\nn\\ 
&= \langle \pi_\Sigma|C_{-1} \bigg(\langle\partial_\Sigma | \xi\rangle
|\Sigma\rangle\bigg) \otimes |V\rangle  +\langle \partial_\xi |
\otimes \langle \pi_\Sigma | \lb \Kmat{13}_0, \Kmat{23}_{-1} \rb
|\xi\rangle \otimes | \Sigma\rangle \otimes |V\rangle 
\;.
\end{align}
We then use \eqref{SeconCom} on the last term
\begin{align} 
& \phantom{=} \langle \partial_\xi | \otimes \langle \pi_\Sigma | \lb
  \Kmat{13}_0, \Kmat{23}_{-1} \rb  |\xi\rangle \otimes | \Sigma\rangle
  \otimes |V\rangle\nn\\ 
&= \langle \partial_\xi | \otimes \langle \pi_\Sigma | \biggl( \lb
  \Kmat{23}_{-1} ,\Kmat{12}_0 \rb  + \Kmat{23}_{-1} \biggr)
  |\xi\rangle \otimes | \Sigma\rangle \otimes |V\rangle \nn\\ 
&= \langle \partial_\xi | \otimes \langle \pi_\Sigma | \biggl(
  \Kmat{23}_{-1} \Kmat{12}_0  - \big(  \Kmat{12}_0
  -1+\overset{12}{\sigma\vphantom{C}} \big)  \,\Kmat{23}_{-1}+
  \overset{12}{\sigma\vphantom{C}}\, \Kmat{23}_{-1} \biggr)
  |\xi\rangle \otimes | \Sigma\rangle \otimes |V\rangle \nn\\ 
&= \langle \partial_\xi | \otimes \langle \pi_\Sigma | \biggl(
  \Kmat{23}_{-1} \Kmat{12}_0 +  \overset{12}{\sigma\vphantom{C}}\,
  \Kmat{23}_{-1} \biggr)  
 |\xi\rangle \otimes | \Sigma\rangle \otimes |V\rangle
 \;,
\end{align}
where we used the section constraint \eqref{SectionConstraint} in the
last step. Together, we obtain  
\begin{align}
\label{eq:mixedcommF}
\lb \mathcal{L}_{\xi,0},\mathcal{L}_{0,\Sigma }\rb  |V\rangle &=
\langle \pi_\Sigma | \, C_{-1}  \bigg( \langle
\partial_\Sigma|\xi\rangle |\Sigma\rangle + \langle \partial_\xi | C_0
|\xi\rangle\otimes |\Sigma\rangle\bigg) \otimes | V\rangle  
\cr
&\qquad + \langle \partial_\xi| C_{-1} ( \langle \pi_\Sigma  |
\xi\rangle |\Sigma\rangle) \otimes |V\rangle 
\;,
\end{align}
which indeed gives a gauge transformation with parameter equal to the
Lie derivative of  
the gauge parameter $| \Sigma \rangle \langle \pi_\Sigma |$:
\begin{align} \cL_\xi \left( | \Sigma \rangle \langle \pi_\Sigma |
  \right) &= \langle \partial_\Sigma|\xi\rangle |\Sigma\rangle \langle
  \pi_\Sigma | + \langle \partial_\xi | C_0( |\xi\rangle\otimes
  |\Sigma\rangle   )\langle \pi_\Sigma |+  |\Sigma\rangle  \langle
  \pi_\Sigma  | \xi\rangle \langle \partial_\xi|  
\  , 
\end{align}
cf., (\ref{LieConstrained}). We recall that the weight of $\ket \Sigma\bra{\pi_\Sigma}$ is shifted due to~\eqref{densityL0}, explaining the absence of the $-1$ in the $C_0$ term in the Lie derivative.

As a last step we consider the commutator of two $\Sigma$ gauge
transformations. Two successive $\Sigma$  transformations give 
\begin{align}
\cL_{0,\Sigma_1} \cL_{0,\Sigma_2} | V \rangle = \langle \pi_{\Sigma_1}
|\otimes \langle \pi_{\Sigma_2} |\, \Kmat{13}_{-1} \,\Kmat{23}{}_{-1}\,
|\Sigma_1\rangle \otimes |\Sigma_2\rangle \otimes |V\rangle 
\;,
\end{align}
so that their commutator is
\begin{align}
\left[ \cL_{0,\Sigma_1},  \cL_{0,\Sigma_2}  \right]| V \rangle &=
\langle \pi_{\Sigma_1} |\otimes \langle \pi_{\Sigma_2} |\lb
\Kmat{13}_{-1}, \Kmat{23}_{-1}\rb |\Sigma_1\rangle \otimes
|\Sigma_2\rangle \otimes |V\rangle\nn\\ 
&= \langle \pi_{\Sigma_1} |\otimes \langle \pi_{\Sigma_2} |\lb
\Kmat{12}_{-1}, \Kmat{13}_{-1}\rb |\Sigma_1\rangle \otimes
|\Sigma_2\rangle \otimes |V\rangle\nn\\ 
&= - \langle \pi_{\Sigma_1} |\otimes \langle \pi_{\Sigma_2} |
\,\Kmat{13}_{-1}  \, \Kmat{12}_{-1} \, |\Sigma_1\rangle \otimes
|\Sigma_2\rangle \otimes |V\rangle\nn\\ 
&=\langle \pi_{\Sigma_1} | \,C_{-1} \left( -\langle \pi_{\Sigma_2} |
\,C_{-1}\, | \Sigma_2\rangle \otimes |\Sigma_1 \rangle\right)  \otimes
| V\rangle \nn\\ 
&= \cL_{0,\tfrac12 ( \langle \pi_{\Sigma_1} |C_{-1} |\Sigma_1\rangle
  \otimes | \Sigma_2\rangle \langle \pi_{\Sigma_2} | - \langle
  \pi_{\Sigma_2} |C_{-1} |\Sigma_2\rangle \otimes | \Sigma_1\rangle
  \langle \pi_{\Sigma_1} | ) }  | V\rangle 
\;,
\end{align}
where we used the identity~\eqref{SeconCom} in the first step, the
section constraint \eqref{SectionConstraintN} in the second, and
finally that the result is antisymmetric, modulo the same section
constraint.  
This concludes the proof of closure of the gauge algebra. 

To summarise, we have shown the closure of transformations (\ref{fullLie})
into an ``algebra''\footnote{As in the lower-dimensional cases, this will not
  be a Lie algebra, since the corresponding brackets do not satisfy
  Jacobi identities. The proper structure, which in the double field
  theory situation is a Courant algebroid, is maybe best described in
  an $L_\infty$ framework \cite{Hohm:2017pnh,BermanCederwallStrickland}.}
\begin{align}
\Big[ \cL_{\xi_1,\Sigma_1},  \cL_{\xi_2,\Sigma_2}  \Big]  =
\cL_{\xi_{12},\Sigma_{12}}
\;,
\end{align}
defined by
\bea
\xi_{12} &\equiv&
\leftbr\xi_1,\xi_2\rightbr\equiv\frac12\left(\LL_{\xi_1}\xi_2-\LL_{\xi_2}\xi_1\right)
\;,\nonumber\\
 | \Sigma_{12} \rangle \langle \pi_{\Sigma_{12}} |&\equiv&
 \cL_{\xi_1} \left( | \Sigma_2 \rangle \langle \pi_{\Sigma_2} | \right) 
  +\frac12 \, \langle \pi_{\Sigma_1} |C_{-1} |\Sigma_2\rangle \otimes | \Sigma_1\rangle \langle \pi_{\Sigma_2} | 
  \nonumber\\
  &&{}
 +  \frac14 \, \langle \partial_{\xi_{2}} | C_1 | \big(| \xi_{2} \rangle \otimes | \xi_1 \rangle - | \xi_{1} \rangle \otimes | \xi_2 \rangle\big)\bra{\partial_{\xi_2}}
 \;\; \;-\;\;( 1 \leftrightarrow 2)
  \;.
  \eea

Finally, it is instructive to decompose the generalised
diffeomorphisms (\ref{fullLie}) under E$_8$, 
and to recover the structure of E$_8$ exceptional field theory.
Expanding the gauge parameter $|\Sigma \rangle \langle  \pi_\Sigma |$
according to (\ref{RL0}) yields 
\bea |\Sigma \rangle \langle  \pi_\Sigma | &=& \left(
 \sigma_1 + \sigma_{2 A} T^A_{-1} + \sigma_{3 AB} T^A_{-1} T^B_{-1} +
 \dots \right) | 0\rangle \langle 0| \nn \\ 
&&\qquad - \left( \Sigma_{0 A}  + \Sigma_{1 A,B} T^B_{-1} + \Sigma_{2
   A,BC} T^B_{-1} T^C_{-1} + \dots \right) | 0\rangle \langle 0| T^A_1  
\;,
\eea
where the indices $AB$ of $ \sigma_{3 AB}$ and the indices $BC$ of
$\Sigma_{2 A,BC} $ are restricted to ${\bf 1} \oplus {\bf 248} \oplus
{\bf 3\,875}$, and similar terms are hidden in the ellipses for all
higher $L_0$ weights. The section constraint implies no constraint on
the parameters $\sigma_{n,\Xi}$, and the parameters $\Sigma_{n,A,\Xi}$
are constrained on their first index 
according to the E$_8$ section constraints (\ref{section3D}).
Similarly, we expand the diffeomorphism parameter $\xi$ as 
\bea | \xi \rangle =\left( \xi^0 + \eta_{AB} \xi_1^A T_{-1}^B+ \xi_{2
  AB} T_{-1}^A T_{-1}^B + \dots \right) |0\rangle \ .  
\label{expXi}
\eea
Assuming partial derivatives of the form (\ref{der3D}), one then obtains
for the Lie derivative 
\bea \cL_{\xi,\Sigma} &=& \xi^0 \partial_0 + \xi_1^A \partial_A -
\partial_A \xi^0 T_1^A + \partial_0 \xi^0 ( L_0-1) + \partial_A
\xi_1^A L_0  
+ \left( f^B{}_{CA} \partial_B \xi_1^C + \Sigma_{0 A}\right) T_0^A 
+ \sigma_1 L_{-1}
\nn \\
&& \ 
 -\left(  \partial_0 \xi_1^A +\Pi^{AB,CD} \partial_B \xi_{2 CD} -
 f^{ABC} \Sigma_{1 B,C}\right)   
 \eta_{AE}T^E_{-1} + \sum_{n > 1} \omega_{n A} T^{A}_{-n} \ , 
 \eea
for some linear combinations $ \omega_{n A}$ of $\partial_0 \xi_{n
  \Xi}, \, \partial_A \xi_{n+1 \Xi},\, \sigma_{n \Xi},\, \Sigma_{n
  A,\Xi}$. It is important to note that, although  $\sigma_{n \Xi}$ is defined in the $L_0$ weight $n-1$ component of $R(\Lambda_0)$, and $\Sigma_{n
  A,\Xi}$ in the tensor product of the $L_0$ weight $n$ component of $R(\Lambda_0)$ with the ${\bf 248}$ of E$_8$, they only appear in $\omega_{n A}$ through an appropriate projection to the ${\bf 248}$ of E$_8$. One understands indeed that $\Sigma$ belongs to the tensor product $R(\Lambda_0)_{-1} \otimes \overline{R(\Lambda_0)} $, but it only appears in the generalised diffeomorphism through a projection to $\mathfrak{e}_9$.

Decomposing the vector $|V\rangle$ accordingly, 
\bea
\ket V =\left( V^0 + \eta_{AB} V_1^A T_{-1}^B+ V_{2
  AB} T_{-1}^A T_{-1}^B + \dots \right) |0\rangle \;,
\eea
  one obtains for the
action on its lowest components 
\bea \cL_{\xi,\Sigma} V^0 &=& \xi^0 \partial_0 V^0 - V^0 \partial_0
\xi^0  + \xi_1^A \partial_A V^0 - V_1^A \partial_A \xi^0 \ ,
\nn\\[1ex] 
 \cL_{\xi,\Sigma} V_1^A &=& \xi^0\, \partial_0 V_1^A - V^0\, \partial_0 \xi_1^A 
  \nn\\&& {}
+ \xi_1^B \partial_B V_1^A + V_1^A \partial_B \xi_1^B 
 - \left( f^{EA}{}_B f^{C}{}_{DE} \,\partial_C \xi_1^D+ f^{CA}{}_B
 \Sigma_{0 C} \right) V_1^B  \nn\\ 
 && \quad - \Pi^{BA,CD} V_{2 CD} \,\partial_B \xi^0 - \Pi^{AB,CD}
 V^0\, \partial_B \xi_{2 CD} +  f^{ABC} \Sigma_{1 B,C} V^0 \ , 
\eea
with $\Pi^{AB,CD}$ from (\ref{Pi}).
The weight of the covariant derivative indicates that in three dimensions, $V^0$ is a
vector field, $V_1^A$ a scalar and $V_{2 AB}$ a 1-form.  
The second line in the Lie derivative of $V_1^A $ reproduces precisely
the E$_8$ internal Lie derivative with respect to the vector field
$\xi_1^A$ and the constrained parameter $\Sigma_{0 A}$
\cite{Hohm:2014fxa}.  
We know from E$_8$ geometry \cite{Hohm:2014fxa,Cederwall:2015ica}
that such a transformation only removes unphysical parts of the vielbein.
In particular, this decomposition illustrates that the additional
gauge transformations in (\ref{fullLie}) 
cannot absorb the standard diffeomorphisms of the first term 
(which ultimately is a consequence of the shift of the $L_0$ charge by
the operator $C_{-1}$). 
The latter thus survive as physical gauge symmetries of the theory as
expected.
Note that the parameters in $\Sigma$ enter in a way that does not
disturb the above interpretation of the transformations of the
lowest components of $\ket V$. This is essential, so that it will not
affect the physical components of a generalised vielbein.

\section{Generalised Scherk--Schwarz reduction} \label{sec:GSS}

We will now perform another consistency check on the proposed form 
of E$_9$ generalised diffeomorphisms (\ref{fullLie}). We will study
the behaviour of these transformations 
under a suitably generalised Scherk--Schwarz Ansatz \cite{Scherk:1979zr} for vectors and
gauge parameters. 
With the internal coordinate dependence of all fields carried by a
Scherk--Schwarz twist matrix $U$ we will 
show that under certain assumptions on this twist matrix, all $Y^M$
dependence in the transformations (\ref{fullLie}) 
consistently factors out such that the generalised diffeomorphisms
translate into an algebraic action on the two-dimensional 
fields. We find that this precisely reproduces the gauge structures
identified in two-dimensional gauged
supergravities~\cite{Samtleben:2007an}.

Before writing down the Scherk--Schwarz Ansatz in the Dirac formalism
we introduce a few definitions. First of all, we need to define the
so-called twist matrix $U$. The group of symmetries of the theory
includes not only E$_9$, but also the Virasoro group ${\rm
  Vir}$~\cite{Julia:1996nu}. In two dimensions, the metric scaling
factor $e^{2\sigma}$ in the conformal gauge $g_{\mu\nu} = e^{2\sigma}
\eta_{\mu\nu}$ scales under the action of the central operator $\dK$  in
$\mathfrak{e}_9$ \cite{Breitenlohner:1986um}. The scalar fields in E$_{8}/($Spin$(16)/
\mathds{Z}_2)$ and their infinite tower of dual scalar fields,
together with the scaling factor $e^{2\sigma}$, parametrise  a coset
element of the central extension of the loop group \cite{Geroch:1970nt,Geroch:1972yt,Breitenlohner:1986um}. On the other hand,
the two-dimensional dilaton $\rho$ is a free field. This field and its
(single) dual $\tilde \rho$ transform non-trivially under the Virasoro
reparametrisations of the loop group spectral parameter $w$. To see
this one observes that an affine redefinition of the spectral
parameter $w\rightarrow a w+b$ can be compensated by the affine
transformation of $(\rho,\tilde \rho)\rightarrow (a \rho , a \tilde
\rho - b)$ \cite{Julia:1996nu}.  These affine transformations define the parabolic
subgroup $\mathds{R}_+ \ltimes \mathds{R}\subset $ SL$(2,\mathds{R})
\subset {\rm Vir}$ generated by $L_0$ and $L_{-1}$.  We therefore
expect that a general Scherk--Schwarz Ansatz will be described by a
twist matrix in the product of this parabolic subgroup and the central
extension of the E$_{8}$ loop group. We decompose the twist matrix
$U$ accordingly as the product of a Virasoro parabolic subgroup
element $U_{\scalebox{0.6}{Vir}}(Y)$ and a loop group element
$U_{\scalebox{0.6}{loop}}(Y)$, which includes both the generators
$T^A_n$ and the central charge generator, 
\begin{align}
\label{Uloop}
 U(Y) = U_{\scalebox{0.6}{Vir}}(Y) U_{\scalebox{0.6}{loop}}(Y) \  .
\end{align}
The definition of the exceptional E$_9$ theory is beyond the scope of
this paper. Nonetheless, we expect that the Scherk--Schwarz Ansatz for
the scalar fields $\mathcal{M}(x,Y)$ and the metric conformal factor $\upsigma(x,Y)$ should be determined
in terms of $U_{\scalebox{0.6}{loop}}(Y)$ as 
\begin{align} 
e^{-2\upsigma(x,Y)} \mathcal{M}(x,Y) = U_{\scalebox{0.6}{loop}}(Y)^T
e^{-2\sigma(x)} M(x)  U_{\scalebox{0.6}{loop}}(Y) \ ,  
\end{align}
whereas the dilaton field and its dual should be determined by\footnote{On the spectral parameter $L_0 = - w \partial_w$ and $L_{-1} = - \partial_w$.}
\begin{align} 
\label{eq:Uvir}
U_{\scalebox{0.6}{Vir}}^{T}(Y)  = e^{\varsigma(Y) L_{-1}}
e^{\upsilon(Y) L_0} \; \Rightarrow \; \uprho(x,Y)  = e^{-\upsilon(Y)}
\rho(x) \; , \quad  \tilde \uprho(x,Y)  = e^{-\upsilon(Y)} (\tilde
\rho(x) - \varsigma(Y)) \; .  
\end{align}
The shift of $\tilde \uprho(x,Y)$ in  $\varsigma(Y)$ is indeed
consistent with the gauging defined in \cite{Samtleben:2007an}, where
the $L_0$ generator is not gauged and so $\upsilon(Y)=0$. Although the
theory remains to be constructed, one can infer from this discussion
that the Scherk--Schwarz Ansatz should involve in general both a
twist matrix in the loop group and a twist matrix in the parabolic
subgroup of $SL(2,\mathds{R})$. Assuming that this is indeed the case,
we shall now see that this permits to define a gauge algebra from the
generalised diffeomorphisms introduced in the last section.  

Note that in higher dimensions one does not only introduce a twist
matrix $U(Y) \in\, $E$_{d}$ (for $d\le 8$), but also a scaling
factor $\rho(Y)$ for the metric field Ansatz, not to be confused with
the dilaton $\uprho(x,Y)$ discussed above. Since the central charge of
the loop algebra acts as a Weyl rescaling of the metric in two
dimensions, this scaling factor $\rho(Y)$ is already included in
$U_{\scalebox{0.6}{loop}}(Y)$ by construction.  

It will be convenient to write the Maurer--Cartan form\footnote{Here,
  we use the notation $U^{-T}\equiv (U^{-1})^T$ to denote the
  transpose of the inverse.} 
\begin{equation} 
\mathcal{J}  = U^{-T} d U^T =  U_{\scalebox{0.6}{Vir}}^{-T}  d
U_{\scalebox{0.6}{Vir}}^{T}  +U_{\scalebox{0.6}{Vir}}^{-T} (
U_{\scalebox{0.6}{loop}}^{-T} d U_{\scalebox{0.6}{loop}}^{T}
)U_{\scalebox{0.6}{Vir}}^{T}  =
\mathcal{J}_{\scalebox{0.6}{Vir}}+\mathcal{J}_{\scalebox{0.6}{loop}}
\ ,  
\end{equation} 
in Dirac notation as
\begin{align}
\label{MC} |\Jmat{} \rangle\langle \Jmat{}| \otimes \langle
\partial_J|  &=   \underline{L}{}_{0} \otimes \langle \partial
\upsilon | + \underline{L}{}_{-1} \otimes e^{-\upsilon} \langle
\partial \varsigma | + \sum_{n} \underline{T}{}_n^A\otimes \langle
j_{n A}|~ +~ \underline{\id} \otimes  \langle j_c| \, , 
\end{align}
where we understand that the $\langle \partial_J|$ bra defines the
derivative index and the $\mathfrak{vir} \oplus \mathfrak{e}_9$ matrix
is written as $|\Jmat{} \rangle\langle \Jmat{}|$. The notation is such
that  
\begin{equation} \label{derTwist} 
 U^T(Y) \otimes \langle \overset{\leftarrow}{\partial}_Y  |  =  U^T
 \underline{L}{}_{0} \otimes \langle \partial \upsilon | +U^T
 \underline{L}{}_{-1} \otimes e^{-\upsilon} \langle \partial \varsigma
 |+ \sum_{n}  U^T  \underline{T}{}_n^A\otimes \langle j_{n A}| +   U^T
 \otimes  \langle j_c|  \ .  
\end{equation}
To distinguish the ket vectors that are acted on the left by $U^T$ and
$U^{-T}$, we use the underlined notation such that in practice, $U^T$
acts on an underlined ket to give a not underlined ket. The same
convention applies to the bra. It follows for instance that the
Maurer--Cartan form \eqref{MC} acts on an underlined ket vector to
give another underlined ket vector, which justifies that we use the
notation $\underline{J}$. The underlined operators are identical to
the non-underlined ones, but are simply understood to act on
underlined kets.

Before spelling out the Scherk--Schwarz Ansatz, it is important to
understand the covariance under the parabolic subgroup $\mathds{R}_+
\ltimes \mathds{R}\subset {\rm Vir}$. The algebraic part of the
generalised Lie derivative (\ref{fullLie}) involves the derivative of the vector field
$|\xi\rangle \langle \partial_\xi |$ through the operator $C_0$, and
the constrained gauge parameter $|\Sigma\rangle \langle \pi_\Sigma|$
through the operator $C_{-1}$. The action of $\mathds{R}_+ \ltimes
\mathds{R}$ on these two operators is determined by the commutation
relation 
\begin{align}
\lb \id \otimes L_m+ L_m\otimes \id, C_n\rb = (m-n)\, C_{m+n} +
\frac{4}{3} m (m^2 -1) \delta_{m+n,0} \, , 
\end{align}
to be such that a twist matrix parametrised as in \eqref{Uloop} acts
on $C_{-1}$ and $C_0$ in the adjoint representation,  
\begin{align} \label{Ctransorm}
\big( U^{T} \otimes U^T \big)\, C_{-1}\,\big( U^{-T} \otimes U^{-T} \big) = e^{\upsilon}
\underline{C}_{-1} \ , \quad \big( U^{T} \otimes U^T \big)\,C_{0}\,\big( U^{-T} \otimes
U^{-T}\big) =\underline{C}{}_0 - \varsigma \underline{C}_{-1} \ ,  
\end{align}
where $\underline{C}_n$ is $C_n$ acting on flattened (underlined)
vectors. Because $|\xi\rangle \langle \partial_\xi |$ and
$|\Sigma\rangle \langle \pi_\Sigma|$ are naturally paired with
$C_{-1}$ and $C_0$, they transform in the coadjoint representation
of the parabolic subgroup $\mathds{R}_+\ltimes \mathds{R}$,
\begin{equation} 
\label{ParaTrans}  |\Sigma  \rangle \langle \pi_\Sigma| \rightarrow
e^{-\upsilon} \bigl( |\Sigma \rangle\langle \pi_\Sigma| + \varsigma
|\xi \rangle   \langle \partial_\xi   | \bigr) \ , \quad  |\xi \rangle
\langle \partial_\xi   | \rightarrow  |\xi \rangle   \langle
\partial_\xi   | \ .  
\end{equation}

The Scherk--Schwarz Ansatz for vectors and gauge parameters written in Dirac notation now takes the form 
\begin{align} \label{SSansatz} 
|V\rangle &=  U^{-T} | \underline{V}\rangle\ , \\
| \xi \rangle &=  U^{-T} |\underline{\xi} \rangle\ , \nn \\
|\Sigma\rangle \langle \pi_\Sigma | &=  e^{-\upsilon} U^{-T} \biggl(
\sum_n  \underline{T}{}_{1+n}^A | \underline{\xi} \rangle\langle j_{n A}
| +  \underline{L}{}_{1} |\underline{\xi} \rangle \langle \partial
\upsilon  |+ \underline{L}{}_{0}  |\underline{\xi} \rangle
e^{-\upsilon} \langle \partial \varsigma  |\biggr) + e^{-\upsilon}
\varsigma | \xi\rangle \langle \partial_\xi | \ , \nn 
\end{align}
where the flat (underlined) ket vectors only depend on external coordinates.
The Ansatz for the vectors $|V\rangle$, $| \xi \rangle$ is of the standard form,
while the Ansatz for the gauge parameter $\Sigma$ matches the $\mathds{R}_+ \ltimes \mathds{R}$  covariance
\eqref{ParaTrans}
 and is explicitly compatible with the constraints (\ref{SectionConstraintSigma})
that this parameter satisfies. Its expression can be written formally for  constant $\varsigma$ in terms of a properly
renormalised trace (see appendix \ref{FormalTrace}) 
\begin{align}
|\Sigma\rangle \langle \pi_\Sigma | &=  e^{-\upsilon}  \frac{1}{\cN}
\Umat{}^{-T} \left( \langle \Jmat{}| \underline{\Kmat{}}_1 |
\Jmat{}\rangle \otimes |\overset{}{\underline{\xi}}\rangle\right)
\langle \partial_J |+ e^{-\upsilon}  \varsigma | \xi\rangle \langle
\partial_\xi | \ ,  
\end{align}  
which exhibits that this Ansatz preserves E$_9$ covariance.

Let us now consider the action of such a generalised diffeomorphism,
\begin{align}
\mathcal{L}_{\xi,\Sigma} |{V} \rangle &= \langle \partial_V|\xi\rangle
\left(  U^{-T} |\underline{V}\rangle\right) + \langle \partial_\xi |
(C_0-1) \left( U^{-T} | \underline{\xi}\rangle \otimes
|V\rangle\right) +  e^{-\upsilon} \varsigma \langle \partial_\xi |
C_{-1} \left( U^{-T} | \underline{\xi}\rangle \otimes |V\rangle\right)
\nn\\ 
&\hspace{-0mm}  +e^{-\upsilon}   \biggl( \langle \partial \upsilon  |
C_{-1} U^{-T} \underline{L}{}_{1} |\underline{\xi}
\rangle+e^{-\upsilon} \langle \partial \varsigma  |  C_{-1} U^{-T}
\underline{L}{}_{0} |\underline{\xi} \rangle  + \sum_n \langle j_{n A}
| C_{-1} U^{-T} \underline{T}{}_{1+n}^A |\underline{\xi}\rangle
\biggr)\otimes  |V\rangle \ ,  
\end{align}
where $\langle \partial_V|$ and $ \langle \partial_\xi |$ are
understood to derive the twist matrix $ U^{-T}$ multiplying
respectively the constant vectors $|\underline{V}\rangle$ and
$|\underline{\xi}\rangle$  using \eqref{derTwist}. Using 
\eqref{Ctransorm} to write everything in terms of flat vectors, one
obtains that the explicit dependence in $\upsilon$ and $\varsigma$ drops
out (such that they only appear through their derivatives $\langle
\partial \upsilon|$ and $e^{-\upsilon} \langle \partial
\varsigma|$). For example  
\begin{align} 
&\phantom{=} -  \langle \underline{j}{}_{nA} | U^T  \otimes U^T  C_0
  U^{-T} \otimes U^{-T}  \underline{T}{}^A_n | \underline{\xi}\rangle+
  e^{-\upsilon} \langle \underline{j}{}_{nA} |  U^T  \otimes U^T
  C_{-1} U^{-T} \otimes U^{-T} ( \underline{T}{}^A_{n+1}- \varsigma
  \underline{T}{}^A_n)  | \underline{\xi}\rangle \nonumber \\ 
&= -\langle \underline{j}{}_{nA} |  ( \underline{C}{}_0 -\varsigma
  \underline{C}{}_{-1}) \underline{T}{}^A_n | \underline{\xi}\rangle+
  e^{-\upsilon} \langle \underline{j}{}_{nA} |  e^{\upsilon}
  \underline{C}{}_{-1}( \underline{T}{}^A_{n+1}- \varsigma
  \underline{T}{}^A_n)  | \underline{\xi}\rangle  \nonumber \\ 
&= -\langle \underline{j}{}_{nA} |   \underline{C}{}_0
  \underline{T}{}^A_n | \underline{\xi}\rangle +\langle
  \underline{j}{}_{nA} |   \underline{C}{}_{-1}
  \underline{T}{}^A_{n+1}  | \underline{\xi}\rangle  \; . 
\end{align}
This exhibits that the Ansatz \eqref{SSansatz} is indeed covariant
with respect to $\mathds{R}_+ \ltimes \mathds{R}$, as advocated
above. For convenience, we introduce the flat derivative bra
$\langle\underline{\partial} | =  \langle\partial | U^{-T}$. One
then obtains 
\begin{align}
U^T  \mathcal{L}_{\xi,\Sigma} |{V} \rangle &= - \Bigl( \langle
\underline{\partial} \upsilon | \underline{\xi} \rangle
\underline{L}{}_0 +e^{-\upsilon}  \langle \underline{\partial}
\varsigma | \underline{\xi} \rangle \underline{L}{}_{-1} + \sum_n
\langle \underline j{}_{nA} | \underline{\xi} \rangle
\underline{T}{}_n^A + \langle \underline{j}{}_c | \underline{\xi}
\rangle \Bigr) | \underline{V} \rangle  \nn \\ 
&\quad -  \Bigl(  \langle \underline{\partial} \upsilon |
(\underline{C}{}_0-1) \underline{L}{}_0 |\underline{\xi} \rangle+
e^{-\upsilon} \langle \underline{\partial} \varsigma  |
(\underline{C}{}_0-1) \underline{L}{}_{-1}  |\underline{\xi} \rangle
\Bigr . \nn\\ 
&\hspace{25mm} \Bigl . + \sum_n \langle \underline j{}_{nA} | (
\underline{C}{}_0-1) \underline{T}{}_n^A | \underline{\xi} \rangle +
\langle \underline{j}{}_c | (\underline{C}{}_0-1)  | \underline{\xi}
\rangle \Bigr)  | \underline{V} \rangle\nn\\ 
&\quad + \Bigl( \langle \underline{\partial} \upsilon |
\underline{C}{}_{-1} \underline{L}{}_1 |\underline{\xi} \rangle  +
e^{-\upsilon}  \langle \underline{\partial} \varsigma |
\underline{C}{}_{-1} \underline{L}{}_0 |\underline{\xi} \rangle  +
\sum_n \langle \underline j{}_{nA} | \underline{C}{}_{-1}
\underline{T}{}_{n+1}^A | \underline{\xi} \rangle  \Bigr)  |
\underline{V} \rangle\nn\\ 
&=  \langle \underline{\partial} \upsilon| \Bigl(
(1-\underline{C}{}_0)   \underline{L}{}_0\otimes \id - \id \otimes
\underline{L}{}_0 + \underline{C}{}_{-1} \,  \underline{L}{}_1\otimes
\id \Bigr) |\underline{\xi} \rangle \otimes  | \underline{V} \rangle-
\langle \underline{j}{}_c | \underline{C}{}_0|\underline{\xi} \rangle
| \underline{V} \rangle  \nn\\ 
&\quad + e^{-\upsilon} \langle \underline{\partial} \varsigma| \Bigl(
(1-\underline{C}{}_0)   \underline{L}{}_{-1}\otimes \id - \id \otimes
\underline{L}{}_{-1} + \underline{C}{}_{-1} \,
\underline{L}{}_0\otimes \id \Bigr) |\underline{\xi} \rangle \otimes
| \underline{V} \rangle \nn\\ 
&\quad + \sum_n \langle \underline{j}{}_{nA} | \Bigl(
(1-\underline{C}{}_0)   \underline{T}{}_n^A \otimes \id - \id \otimes
\underline{T}{}_n^A + \underline{C}{}_{-1} \,  \underline{T}{}_{n+1}^A
\otimes \id \Bigr) |\underline{\xi} \rangle \otimes  | \underline{V}
\rangle\nn\\ 
&= - \Bigl( \langle \underline{\partial} \upsilon|
(\underline{L}{}_0+1) + e^{-\upsilon} \langle  \underline{\partial}
\varsigma| \underline{L}{}_{-1} + \sum_n \langle \underline{j}{}_{nA}
| \underline{T}{}_n^A+  \langle \underline{j}{}_c |\Bigr)
\underline{C}{}_0  |\underline{\xi} \rangle \otimes  | \underline{V}
\rangle\nn\\ 
&\hspace{15mm} + \Bigl( \langle \underline{\partial} \upsilon|
\underline{L}{}_1 +e^{-\upsilon}\langle \underline{\partial}
\varsigma| (\underline{L}{}_0-1) + \sum_n \langle \underline{j}{}_{nA}
| \underline{T}{}_{n+1}^A \Bigr) \underline{C}{}_{-1}
|\underline{\xi} \rangle \otimes  | \underline{V} \rangle  
\; ,
\label{actionxiV}
\end{align}
where in the last step we have used
\begin{align} 
[ C_0 , L_n \otimes \id ] - L_n \otimes \id + \id \otimes L_n &= [
  C_{-1} , L_{n+1} \otimes \id ]  + C_n \ ,\nn\\ 
[ C_0 , T_n^A \otimes \id ] - T_n^A \otimes \id + \id \otimes T_n^A &=
[ C_{-1},T_{n+1}^A \otimes \id] \ ,  
\end{align}
which one computes straightforwardly using the definition of $C_n$.  
Defining
\begin{align} 
\langle \underline{\theta}|  &\equiv \langle \underline{\partial}
\upsilon | \underline{L}{}_1 +e^{-\upsilon} \langle
\underline{\partial} \varsigma| (\underline{L}{}_0-1) + \sum_n \langle
\underline{j}{}_{nA}| \underline{T}{}^A_{n+1} \ , \nonumber \\   
 \langle \underline{\vartheta}|  &\equiv -  \langle
 \underline{\partial} \upsilon | (\underline{L}{}_0+1) - e^{-\upsilon}
 \langle  \underline{\partial} \varsigma| \underline{L}{}_{-1}  -
 \sum_n \langle \underline{j}{}_{nA}| \underline{T}{}^A_n - \langle
 \underline{j}{}_c| \  , 
\label{embedding}
\end{align}
 the action (\ref{actionxiV}) can be put in the compact form
\begin{align}
\delta_{\underline{\xi}}   | \underline{V} \rangle \equiv U^T
\mathcal{L}_{\xi,\Sigma} |{V} \rangle = 
 \langle \underline{\theta}| \,\underline{C}{}_{-1}\, |
 \underline{\xi}\rangle\otimes |\underline{V}\rangle +  \langle
 \underline{\vartheta} | \,\underline{C}{}_0\,| \underline{\xi}\rangle
 \otimes |\underline{V}\rangle 
\;.
\label{2Dgauge}
\end{align}
A consistent reduction thus corresponds to a twist matrix constructed
such that the combinations (\ref{embedding}) 
are constant, corresponding to two different types of gaugings.
The first one, parametrised by a constant embedding tensor $\langle \underline{\theta}|$,
precisely reproduces the standard gauge structure of
two-dimensional gauged supergravity \cite{Samtleben:2007an}. 
The second type of gauging, parametrised by a constant $\langle \underline{\vartheta}|$, is slightly less standard. 
As follows from (\ref{2Dgauge}), it gauges the generator $\dL \in \mathfrak{e}_9$ that is represented as $L_0$,
which is not a symmetry of the ungauged Lagrangian. The resulting gaugings thus do not admit an action but are
defined only on the level of their field equations. In this sense they are the analogues
of the trombone gaugings \cite{LeDiffon:2008sh} 
that gauge the trombone scaling symmetry \cite{Cremmer:1997xj}
of higher-dimensional supergravities. Note that in the two-dimensional case the trombone symmetry as defined in
 \cite{Cremmer:1997xj} is an ordinary (and off-shell)  
 Weyl symmetry of the two-dimensional  theory that is generated by the
 central charge $\dK$ of $E_9$.  
 It is gauged by both parameters 
 $\langle \underline{\theta}|$ and $\langle \underline{\vartheta}|$
 and thus part of a generic gauging in two dimensions.

Explicitly, one has
\bea &\phantom{=}& \langle \underline{\theta}| \underline{C}{}_{-1} |
\underline{\xi}\rangle  +  \langle \underline{\vartheta} |
\underline{C}{}_0| \underline{\xi}\rangle  \\ 
 &=& \bigl(  \langle \underline{\theta}| \underline{L}{}_{-1}|
\underline{\xi}\rangle + \langle \underline{\vartheta}|
\underline{L}{}_0 | \underline{\xi}\rangle \bigr) - \eta_{AB}
\sum_{n\in\ZZ} \bigl(  \langle \underline{\theta}|
\underline{T}{}_{-n-1}^A | \underline{\xi}\rangle+ \langle
\underline{\vartheta}| \underline{T}{}_{-n}^A | \underline{\xi}\rangle
\bigr)   \underline{T}{}_{n}^B  + \langle \underline{\theta}|
\underline{\xi}\rangle \underline{L}{}_{-1} +  \langle
\underline{\vartheta}| \underline{\xi}\rangle \underline{L}{}_{0} \nn 
 \ .\eea

A straightforward computation shows that the algebra of gauge
transformations (\ref{2Dgauge})  
closes according to
\bea
\lb \delta_{\xi_1},  \delta_{\xi_2} \rb |\underline{V}\rangle &=&
\delta_{\xi_{12}}\,|\underline{V}\rangle
\;,
\eea
with gauge parameter
\bea
|\xi_{12}\rangle &\equiv&
\frac12 \bigl( 
  \langle \underline{\theta}| \underline{C}_{-1} 
  +  \langle \underline{\vartheta} | \underline{C}_0\bigr)  \bigl(
  | \underline{\xi}_{1}\rangle \otimes  |\underline{\xi}_2\rangle-  |
  \underline{\xi}_{2}\rangle  \otimes |\underline{\xi}_1\rangle \bigr)  \;, 
\eea
provided that the components of the embedding tensor satisfy the constraints
\begin{align}
 \langle \underline{\theta}| \otimes \langle \underline{\theta}|  {\underline{C}}_{-1}
+  \langle \underline{\vartheta} |\otimes \langle \underline{\theta}| \left(
{\underline{C}}_{0} 
+{\sigma} 
- 1 \right) &=0
\;,
\nonumber\\
   \langle \underline{\vartheta} | \otimes  \langle \underline{\vartheta}
   |{\underline{C}}_{0}  
  +  \langle \underline{\theta}| \otimes \langle \underline{\vartheta} |
  {\underline{C}}_{-1}&=0 
\;.
\end{align}
If the twist matrix from which this embedding tensor 
is obtained satisfies the section constraint, these constraints must be
automatically satisfied since closure of the algebra is guaranteed by construction by the closure of the generalised diffeomorphism algebra.
In the absence of an $L_0$-gauging ($\langle \underline{\vartheta}|=0$),
we recover the condition 
\bea
\langle \underline{\theta}| \otimes \langle \underline{\theta}|   \,{\underline{C}}_{-1} =0
\;,
\eea
which had been identified as the quadratic constraint on the embedding tensor 
in \cite{Samtleben:2007an}. For pure $L_0$-gaugings ($\langle
\underline{\theta}|=0$) on the other hand, 
we precisely recover the section constraint
\bea
 \langle \underline{\vartheta} | \otimes \langle \underline{\vartheta} | \,
 {\underline{C}}_{0} =0 
\; , 
\eea
as for pure trombone gaugings in higher dimensions.

\section{Generalisation to other groups\label{GeneralisationSection}}

In this section, we discuss two generalisations of our formul\ae{} for the generalised diffeomorphisms~\eqref{fullLie0} and section constraint~\eqref{SectionConstraint0}. The first generalisation is to arbitrary affine algebras and the second one to arbitrary Kac--Moody algebras. 
In the most general case, the generalisations we present only give the
generalised form of the section constraint and generalised Lie
derivative, but we have not checked directly closure of the gauge
algebra which also requires the introduction of extra constrained
transformation parameters $\Sigma$. For the generalisation to other
affine algebras with coordinates in the basic representation, the
parameter $\Sigma$ can be defined in analogy with the $\mf{e}_9$ case
considered in detail above and the gauge algebra closes in exactly the
same way. For general Kac--Moody algebras, a systematic introduction
of $\Sigma$  most probably requires the language and properties of
tensor hierarchy algebras that we shall not attempt here. We also note
that even if a consistent gauge algebra is established, this does not
guarantee the existence of a non-trivial physical model for any
Kac--Moody algebra. 

\subsection{Extension to other affine groups\label{AffineSection}}

In this section, we discuss how much of the structure of the E$_9$
exceptional geometry will carry over to affine extensions of
other ``exceptional'' field
theories based on simple symmetry groups in $D=3$ space-time
dimensions~\cite{Breitenlohner:1987dg,Cremmer:1999du}.\footnote{The
  case of semi-simple symmetries and their affine and further
  extensions was discussed in~\cite{Kleinschmidt:2008jj}.} An example
of a double field theory with SO$(8,n)$ symmetry with three external
dimensions was recently constructed in~\cite{Hohm:2017wtr}, the
duality covariant  
theory based on the Ehlers group SL$(2)$ was constructed
in~\cite{Hohm:2013jma}, and the picture for higher SL$(n)$ was given
in \cite{Cederwall:2015ica}.

The important steps in the construction of the E$_9$ exceptional
geometry performed in this paper were $(i)$ the identification of an
appropriate representation $R(\Lambda_0)$ for the coordinates,
$(ii)$ the identification of an appropriate section constraint in
$\overline{R(\Lambda_0)}\otimes \overline{R(\Lambda_0)}$ and $(iii)$
verification of the closure of the generalised diffeomorphisms up to
section constraint. It is noticeable that in the definition of the
generalised diffeomorphism~\eqref{fullLie} and section
constraint~\eqref{SCall} only the coset Virasoro generators
appear. Little use of the structure of E$_8$ itself is made.   

Let us consider an arbitrary simple finite-dimensional algebra
$\mf{g}$ (replacing $\mf{e}_8$) and its associated (non-twisted)
affine extension $\mf{g}^+$ (replacing $\mf{e}_9$). The associated
groups will be denoted by $G$ and $G^+$, respectively. The known
structure of exceptional field theory with $G$ symmetry have internal
coordinates in the adjoint representation $\textrm{{\bf adj}}$ of $G$
satisfying a section constraint in the representation $\textrm{\bf
  sec}$ of $G$ that lies in the tensor product of two adjoint
representations. The pieces of the section constraint that lie in the
symmetric part of the tensor product correspond to the
three-dimensional embedding tensor
(as a consequence of the duality between level 2 and level $-1$ in the
tensor hierarchy algebra for compactifications to three dimensions \cite{Palmkvist:2013vya}).
There is also an anti-symmetric
contribution to the section
constraint~\cite{Hohm:2013jma,Hohm:2014fxa,Cederwall:2015ica,Hohm:2017wtr}.
In addition, the generalised
Lie derivatives with three external dimensions contain also constrained
parameters $\Sigma$ besides the standard parameters $\xi$. The
standard physical solution to the $D=3$ section constraint is given by
taking from $\textrm{\bf adj}$ a $d$-dimensional subspace that
corresponds to the maximal number of dimensions that can be
oxidised~\cite{Cremmer:1999du,Keurentjes:2002xc}. 

All affine algebras afford a ``basic'' representation $R(\Lambda_0)$ at
level $k=1$~\cite{Goddard:1986bp,Kac:1990gs}. Its distinguishing
property is that it is an irreducible highest weight module of
$\mf{g}^+$ that decomposes under $\mf{g}$ as 
\begin{align}
R(\Lambda_0) = {\bf 1}_0 \oplus \textrm{{\bf adj}}_{-1} \oplus
\textrm{\bf sec}_{-2} \oplus\ldots\,,
\end{align}
where the antisymmetric part of {\bf sec} (in the tensor product of
two adjoints) is $\textrm{\bf sec}_a=\textrm{\bf adj}$.
This is the generalisation of~\eqref{RL0}.
It is a generic property of $R(\Lambda_0)$ that there are null
  states at affine level $-2$. They are a consequence of
  $f_0f_0|0\rangle=0$, where $f_0$ is the generator corresponding to
  the root $-\alpha_0$. It is easily shown that this state is
  annihilated by $e_0$. In terms of $\mf{g}$, the state $f_0f_0|0\rangle$ would
  carry the weight $2\theta$, where $\theta$ is the highest
root of $\mf{g}$. Therefore, the ``big'' representation in the
symmetric product of two $\mf{g}$-adjoints is always absent at affine
level $-2$ in $R(\Lambda_0)$, and the symmetric part of {\bf sec} is
some smaller representation:
$\textrm{\bf adj}\otimes_s\textrm{\bf adj}=r(2\theta)\oplus\textrm{\bf sec}_s$.

We can then work out the general tensor product of two elements in
$R(\Lambda_0)$ at
low $\mf{g}$ levels, 
\begin{align}
\label{R0R0}
R(\Lambda_0)\otimes R(\Lambda_0) = {\bf 1}_0
\oplus \left(2\cdot \textrm{\bf adj}\right)_{-1} \oplus \left( 2\cdot
\textrm{\bf sec} \oplus \textrm{\bf adj} \otimes\textrm{\bf
  adj}\right)_{-2} \oplus \ldots\,. 
\end{align}
A physically expected solution to the $D=2$ section constraint is taking
the singlet at level $0$ and the $d$-dimensional subspace in
$\textrm{\bf adj}$ that corresponds to the solution of the $D=3$ section
constraint. These $d+1$ coordinates together correspond to the
oxidation from $D=2$ external space to the same maximal oxidation
endpoint in $3+d$ dimensions. We would therefore like the $D=2$ section
constraint to be strong enough to remove everything but a solution of
this type. 

{}From the representation theory of affine algebras we know that the
tensor product $R(\Lambda_0)\otimes R(\Lambda_0)$ decomposes into
representations at level 
$k=2$. More precisely, there is again a coset construction similar
to~\eqref{TensorTwoROne} where the standard modules at $k=2$ appear
multiplied with coset Virasoro characters that are $q$-series. What
types of $k=2$ modules exist does depend on the structure of
$\mf{g}^+$. Two $k=2$ modules that always exist are 
\begin{align}
R(2\Lambda_0) = {\bf 1}_0 \oplus \textrm{\bf adj}_{-1}
\oplus(\textrm{\bf sec}\oplus r(2\theta))_{-2}\oplus\ldots
\end{align}
that is the leading term in the symmetric part of the tensor product and
\begin{align}
  R(\Lambda_1)_{-1} =  \textrm{\bf adj}_{-1} \oplus
  (\textrm{\bf adj}\wedge\textrm{\bf adj}\oplus\textrm{\bf
    sec}_s)_{-2}
  \oplus\ldots
\end{align}
by which we denote the leading term in the anti-symmetric part of the
tensor product\footnote{The notation may seem to indicate that there is a
unique simple root $\alpha_1$ connected with a single line to
$\alpha_0$. This is not necessarily the case (\eg\ in $A_n^+$);
then $\Lambda_1$ has
to be reinterpreted as the weight
$\sum_{i=1}^{\textrm{rank}\,\fg}a_i\Lambda_i$, where the 
highest root of $\fg$ is $\theta=\sum_{i=1}^{\textrm{rank}\,\fg}a_i\lambda_i$.}.
The first null states in $R(2\Lambda_0)$ appear at level $-3$.
The null states in $r(2\theta)$
  at affine level $-1$ in $R(\Lambda_1)$ come
from the observation that $f_0|\Lambda_1\rangle$ is a null state.

Comparing these two leading expansions with the full tensor
product~\eqref{R0R0} we conclude that any other $k=2$ module can only
start contributing from level $-2$ onwards. Writing the levels as a
$q$-series this means that 
\begin{align}
\label{eq:R0R0q}
R(\Lambda_0)\otimes_s R(\Lambda_0)
&=(1+q^2)R(2\Lambda_0)
\oplus q^2\textrm{\bf sec}'_s\oplus\ldots\nn\\
R(\Lambda_0)\wedge R(\Lambda_0)
&=(q+q^2)R(\Lambda_1)\oplus\ldots
\end{align}
where the ellipses denote terms at affine level $-3$ and lower, and where
$\textrm{\bf sec}_s={\bf1}\oplus\textrm{\bf sec}'_s$. Note that the
identification of an irreducible highest weight affine representation
from its leading irreducible $\fg$ representation is unique at a given $k$. 
What is noteworthy is the absence of
a term at level $-1$ in the $R(2\Lambda_0)$ piece.

There is a coset Virasoro construction associated with the tensor
product of two $k=1$ modules. The $q$-series (after an appropriate
shift of the conformal weight of the affine representations) are
characters of this coset Virasoro algebra. Unlike the case for E$_9$,
it is not true in general that they are characters in the minimal
series since the central charge can be $c \geq
1$~\cite{KacWakimoto1988}.\footnote{Another case where one has
  $c=\frac12$ as for E$_9$ is the affine extension $A_1^+$  of
  SL$_2(\mathds{R})$ (the Geroch group
  \cite{Geroch:1970nt,Geroch:1972yt,Breitenlohner:1986um}) corresponding to pure
  four-dimensional Einstein 
  gravity. Also the coset constructions based on A$_2$ or any
  finite-dimensional exceptional algebra fall in the minimal series.}
Nevertheless, the $q$-series always represent characters
of (possibly reducible) unitary representations of the Virasoro
algebra.  

The contribution to $h$ from the $\fg$ quadratic Casimir is 0 for
$R(2\Lambda_0)$ and $\frac{g^\vee}{2+g^\vee}$ for $R(\Lambda_1)$. The
generalisation of (\ref{TensorTwoROne}) is
\begin{align}
R(\Lambda_0)\otimes R(\Lambda_0)
=\Vir_0\otimes R(2\Lambda_0)_0
\oplus\Vir_{-\frac{2}{2+g^\vee}}\otimes R(\Lambda_1)_{-\frac{g^\vee}{2+g^\vee}}
\oplus\ldots 
\label{TensorTwoROneGeneral}
\end{align}
where the subscript on the coset Virasoro modules is $-h$.
If we define the rescaled coset Virasoro operators
$C_n=(2+g^\vee)L_n^{\rm coset}$, we can conclude that the appropriate
section constraints remain of the precise form (\ref{SCall}):
\begin{align}
  \langle \partial_1 | \otimes
  \langle \partial_2 |  \, 
 ( C_0 - 1 + \sigma )&=0 \ ,\nn \\
  \langle \partial_1 | \otimes \langle
 \partial_2 |   \, C_{-n} &=0 \;,\quad \forall n>0\;, \\ 
 \left(  \langle \partial_1 | \otimes
 \langle \partial_2 |  +  \langle \partial_2 | \otimes \langle
 \partial_1 |\right) \,C_1 &=0 \;, \nn
 \end{align}
since
$C_0$ then takes the value $0$ and $2$ in the leading symmetric and
antisymmetric states, respectively.
Then, the only modules remaining in the product of two derivatives
are the leading ones, corresponding to the highest weights in the
(conjugate) Virasoro modules corresponding to $\overline{R(2\Lambda_0)}$ and
$\overline{R(\Lambda_1)}$.

Since all the remaining steps in the calculation only depend on the
coset Virasoro algebra, we conclude that the form of the generalised
diffeomorphism and the closure of the gauge algebra proceed in the
same way for all affine symmetries $G^+$.

\subsection{Strong section constraint for an arbitrary Kac--Moody
  algebra
\label{ArbitraryKMSection}}

The section constraint is an important starting point for the
construction of any ``extended geometry'', be it double or exceptional
field theory, or some other model with enhanced symmetry algebra $\fg$.
The actual form of the $Y$ tensor defining this constraint has normally been determined on a
case-by-case basis. This applies in particular to exceptional field
theory, where it is notoriously difficult to find tensorial identities
applying to every member of the series of exceptional algebras.
However, in \cite{Palmkvist:2015dea} a general construction of the $Y$ tensor 
was given,
based on bosonic and fermionic extensions of the algebra $\fg$.
The identities needed for closure and covariance of the generalised Lie derivative
are then automatically
satisfied, except for one of them (whose failure is the reason for introducing an extra constrained transformation parameter
in the $\mathfrak{e}_8$ case). 

The construction of the $Y$ tensor in \cite{Palmkvist:2015dea} was given explicitly for exceptional field theory, but can easily be generalised to
any highest (or lowest) weight representation $R(\lambda)$ of
any Kac--Moody algebra $\mathfrak{g}$, except for cases where $\fg$ or its fermionic extension
has a degenerate Cartan matrix. In this section we will obtain a general formula 
for the $Y$ tensor which includes also the degenerate cases, and thus 
encompasses
all the known finite-dimensional examples and the affine algebra
examples described in this paper.
We will restrict to simply laced
$\fg$. In general we consider the Lie group $G$ defined in
\cite{Peterson:1983} for an arbitrary Kac--Moody algebra
$\mathfrak{g}$.  

A vector $\ket p$ in a highest weight
representation $R(\lambda) $ satisfies
the (weak) section constraint if $\ket p\otimes\ket p\in
R(2\lambda)$. This is equivalent to the statement that $\ket p$ is in
a minimal $R(\lambda)$-orbit under $\fg$. This is discussed 
\eg\ in \cite{Berman:2012vc}, and a direct connection
between minimal orbits
and Borcherds superalgebras (the fermionic extensions of $\fg$) was made in \cite{Cederwall:2015oua}.

The quadratic Casimir,
\bea
C_2=\frac12\eta_{AB} :T^AT^B\hspace{-1mm} :\hspace{2mm} =\sum_{\alpha\in \Delta_+}E_{-\alpha}E_\alpha+\frac12(H,H)
+(\varrho,H)\; , 
\eea
is defined for finite- and infinite-dimensional Kac--Moody algebras on a highest
weight module, where the Weyl vector $\varrho$ is the sum of the fundamental
weights (instead of half the sum of the possibly infinitely many positive roots
in $\Delta_+$). It is normalised by
$C_2(R(\lambda))=\frac12(\lambda,\lambda+2\varrho)$, so that
$C_2(\hbox{adj})=g^\vee$ for finite-dimensional $\fg$. Here $T^A$ are
the generators of $\fg$ and $\eta_{AB}$ is the invariant symmetric bilinear form. 
The last term is a normal ordering term, which for finite-dimensional
$\fg$ can be absorbed into a symmetrically ordered product of
generators.
We observe that
$C_2(R(2\lambda))=2C_2(R(\lambda))+(\lambda,\lambda)$. Also,
there is no other irreducible highest weight representation in the
symmetric product of $R(\lambda)$ with itself with this maximal value
of $C_2$.

The weak section constraint on $\ket p$ is equivalent to the equation
\begin{align} \label{SimSec} 
0&=\left[C_2(R(2\lambda))-2C_2(R(\lambda))-(\lambda,\lambda)\right]
\ket p\otimes\ket p\nn\\
&=\frac12\eta_{AB} :T^AT^B: (\ket p\otimes\ket p)
-\left (\frac12\eta_{AB} :T^AT^B: \ket p \right)\otimes\ket p\nn\\
&\qquad-\ket p\otimes\left( \frac12\eta_{AB} :T^AT^B: \ket p \right) - (\lambda,\lambda)\ket
p\otimes\ket p\nn\\
&=\left[\eta_{AB}T^A\otimes T^B-(\lambda,\lambda)\right]\ket
p\otimes\ket p
\;.
\end{align}
Any vector satisfying $\ket p\otimes\ket p\in R(2\lambda)$ satisfies this equation by construction and it was proven in \cite{Peterson:1983} that all the solutions to this equation are in the $\mathds{R}^\times \times G$-orbit of the highest weight vector $\ket \lambda$ of $R(\lambda)$.\footnote{The rescaling factor $\mathds{R}^\times$ is not included in $G$ when $(\lambda,\lambda)=0$, unless $\mathfrak{g}$ includes a central charge. This would be the case for $E_{10}$ for example.} This equation determines therefore the unique minimal non-trivial $G$-orbit in $R(\lambda)$, where the $\mathds{R}^\times$ is related to rescalings in the one-dimensional highest weight space.

In order to define the strong section constraint we now consider a second vector $\ket q$ such that all $\ket p$, $\ket q$ and $\ket p+\ket q$ satisfy the section constraint. Since \eqref{SimSec} is by construction $G$ invariant, one can assume without loss of generality that $\ket p = \ket \lambda$, the highest weight vector. Then it is convenient to decompose 
\bea 
\label{eq:qdec}
\ket q = \sum_{k=0}^n \ket q_k \ , \qquad (\lambda , H) \ket q_k = ( ( \lambda, \lambda) - k) \ket q_k \ , 
\eea
and the positive roots as $\Delta_+ = \sum_{k\ge 0} \Delta_k$ such that $\alpha_k\in \Delta_k$ satisfies $( \lambda, \alpha_k) = k$, and  $n$ is the lowest weight for which $\ket q_n$ is non-zero. Because the weight is preserved by the operator $\eta_{AB}T^A\otimes T^B$, one obtains that the lowest weight component of the constraint on $\ket p\otimes \ket q+\ket q\otimes \ket p$ reduces to
\begin{align} 
0&= \Bigl( \left[\eta_{AB}T^A\otimes T^B-(\lambda,\lambda)\right]( \ket
\lambda\otimes\ket q+\ket q\otimes\ket \lambda)  \Bigr)_n  \\ 
&= \left[\eta_{AB}T^A\otimes T^B-(\lambda,\lambda)\right]( \ket
\lambda\otimes\ket q_n+\ket q_n\otimes\ket \lambda)  \nn \\
&= - n \ket \lambda \otimes \ket q_n -n \ket q_n \otimes \ket \lambda + \sum_{k=1}^n \sum_{\alpha_k\in \Delta_k} \bigl(  E_{-\alpha_k} \ket  \lambda \otimes E_{\alpha_k} \ket q_n+E_{\alpha_k} \ket q_n \otimes E_{-\alpha_k} \ket  \lambda  \bigr) \nn \ ,   
\end{align}
which in turn can only be satisfied if 
\bea  
n \ket \lambda \otimes \ket q_n  = \sum_{\alpha_n \in \Delta_n } E_{\alpha_n} \ket q_n \otimes E_{-\alpha_n} \ket  \lambda \ . 
\eea
The only solution is 
\bea  
\ket q_n  =  \sum_{\alpha_n \in \Delta_n }  v_{\alpha_n} E_{-\alpha_n}  \ket  \lambda \ . 
\eea
Recalling from~\eqref{eq:qdec} that $n$ is the maximal value for which $\ket q_n$ is non-zero, we now consider the lowest weight component of the constraint on $\ket q\otimes \ket q$, {\it i.e.} 
\bea 
&& \Bigl( \left[\eta_{AB}T^A\otimes T^B-(\lambda,\lambda)\right] \ket
q\otimes\ket q   \Bigr)_{2n}  \\ 
&=& \sum_{\alpha_n, \beta_n \in \Delta_n} v_{\alpha_n} v_{\beta_n}  \left[\eta_{AB}T^A\otimes T^B-(\lambda,\lambda)\right]  E_{-\alpha_n}  \ket  \lambda \otimes E_{-\beta_n}  \ket  \lambda \nn \\ 
&=&  \hspace{-4mm} \sum_{\alpha_n, \beta_n \in \Delta_n} \hspace{-3mm}  v_{\alpha_n} v_{\beta_n}  \Bigl( \bigl(( \alpha_n , \beta_n) - 2  n \bigr) E_{-\alpha_n}  \ket  \lambda \otimes E_{-\beta_n}  \ket  \lambda +  \sum_{ \gamma \in \Delta} [ E_{\gamma} ,E_{-\alpha_n}  ] \ket  \lambda \otimes [ E_{-\gamma} ,  E_{-\beta_n}]   \ket  \lambda  \Bigr) \nn \ .   \eea
There is a lowest weight $\lambda-\alpha_n$ such that $v_{\alpha_n}\neq 0$, {\it i.e.}, $v_{\alpha_n+\gamma_0}=0$ for all positive $\gamma_0$ on level $0$. This implies that
there is no contribution to the term in $ (( \alpha_n ,
\alpha_n) - 2  n ) E_{-\alpha_n}  \ket  \lambda \otimes E_{-\alpha_n}
\ket  \lambda$ from $[ E_{\gamma_0}, E_{-\alpha_n-\gamma_0}]\ket
\lambda\otimes [ E_{-\gamma_0}, E_{-\alpha_n+\gamma_0}]\ket \lambda$
and  we must therefore have $(\alpha_n,\alpha_n) = 2 n$.
Since in general
$(\alpha_n,\alpha_n)\le2$ for any Kac--Moody algebra, the constraint on $\ket q$ can only have
solutions with $n=1$. 

We thus have 
\bea 
\label{eq:qrep}
\ket q  = \Bigl( v_0 + \sum_{\alpha_1 \in \Delta_1} v_{\alpha_1} E_{-\alpha_1} \Bigr) \ket \lambda \ , 
\eea
and the weak section constraint reduces to
\bea && \left[\eta_{AB}T^A\otimes T^B-(\lambda,\lambda)\right] \ket
q\otimes\ket q   \nn \\ 
&=&  \hspace{-4mm} \sum_{\alpha_1, \beta_1 \in \Delta_1} \hspace{-3mm}  v_{\alpha_1} v_{\beta_1}  \biggl( \bigl(( \alpha_1 , \beta_1) - 2  \bigr) E_{-\alpha_1}  \ket  \lambda \otimes E_{-\beta_1}  \ket  \lambda +\hspace{-2mm}   \sum_{\pm  \gamma_0 \in \Delta_0} [ E_{\gamma_0} ,E_{-\alpha_n}  ] \ket  \lambda \otimes [ E_{-\gamma_0} ,  E_{-\beta_n}]   \ket  \lambda  \biggr .  \nn\\
&& \biggl . \hspace{50mm} + E_{-\alpha_1} E_{-\beta_1}  \ket  \lambda \otimes \ket \lambda + \ket \lambda \otimes E_{-\alpha_1} E_{-\beta_1}  \ket  \lambda \biggr) \ .   \eea
The vector $\ket q$ automatically solves the section constraint if $v_{\alpha_1}$
is only non-zero for the simple root dual to $\lambda$, and by
construction for any $v_{\alpha_1}$ obtained from the latter by the
action of the stabiliser $G_0$ of $(\lambda,H)$. The same theorem from \cite{Peterson:1983} implies then moreover that all the solutions are $G_0$-conjugate to this one.

Now we can compute for the orbit representative $\ket q\otimes \ket\lambda$ (with $\ket q$ as in~\eqref{eq:qrep}) that
\bea 
&& \left[\eta_{AB}T^A\otimes T^B-(\lambda,\lambda)\right] \ket
q \otimes\ket\lambda   \nn \\ 
&=& \sum_{\alpha_1 \in \Delta_1} v_{\alpha_1} \Bigl( - E_{-\alpha_1} \ket \lambda \otimes \ket \lambda + \sum_{\beta_1\in \Delta_1} E_{\beta_1} E_{-\alpha_1} \ket \lambda \otimes E_{-\beta_1} \ket \lambda \Bigr) \nn\\ 
&=& - \ket q \otimes \ket \lambda + \ket \lambda \otimes \ket q  \ .   
\eea 
By $G$-covariance we therefore have that the strong section constraint on any pair of vectors $\ket p$ and $\ket q$ therefore implies in general that $Y \ket p\otimes \ket q = 0$ for the tensor 
\bea 
\sigma Y = -\eta_{AB}T^A\otimes T^B+(\lambda,\lambda)+\sigma-1 \ .\label{YTevvvvnsor}  
\eea

This tensor permits to define the generalised diffeomorphisms uniquely and uniformly for any group $G$ and highest weight representation $R(\lambda)$ as
\begin{align} 
\label{eq:GenLie}
\mathcal{L}_{\xi} |{V}\rangle &=
 \langle{\partial}{}_V|{\xi}\rangle |{V}\rangle  -  \langle{\partial}{}_\xi |V \rangle |{\xi}\rangle + \langle \partial_\xi |Ê\sigma Y |\xi\rangle \otimes | V\rangle \nn\\
 &=  \langle{\partial}{}_V|{\xi}\rangle |{V}\rangle  -
 \eta_{AB}\langle \partial_\xi |T^A |\xi\rangle T^B  | V\rangle +
 \bigl( ( \lambda,\lambda) - 1 \bigr) \langle \partial_\xi |\xi\rangle
 | V\rangle \ ,   
\end{align}
such that they reduce to standard diffeomorphisms if $|V\rangle$ and
$\ket \xi$ satisfy the strong section constraint, and the
connection term is valued in $ \mathds{R} \oplus \fg$. The  
$Z$ tensor is then as usual $Z=Y-1$. The overall normalisation of the
$Y$ tensor is of course not determined by the homogeneous condition $Y\ket
p\otimes\ket q=0$, but follows from demanding that\footnote{In the
  affine case the scaling is included in $\fg$ through the central extension.}
$\sigma Z\in(\fg\oplus\mathds{R})\otimes(\fg\oplus\mathds{R})$, \ie,
that the term $\sigma$ in (\ref{YTevvvvnsor})
cancels in $\sigma Z=\sigma Y-\sigma$.
Note, however, that the closure of these candidate generalised diffeomorphisms
is not guaranteed by the construction, neither is it expected that any
choice of algebra and representation will lead to a meaningful
field theory.

The remarkably simple expression~\eqref{eq:GenLie} turns out to reproduce (by necessity) the
invariant tensors used in all previously constructed extended
geometries. In particular $(\lambda,\lambda) -1  = \frac{1}{9-d}$ for
E$_{d}$ type groups with $3 \le d \le 11$, except for $d=9$, in which
case one gets instead $(\lambda,\lambda)=0 $ as described in this
paper. They also generalise to arbitrary Kac--Moody algebras, and have
therefore a potential to be applicable also \eg\ in E${}_d$, $d>9$. 

We also remark that the construction above can be used to recover ordinary Riemannian geometry as well by taking $\mf{g}=\mf{sl}(n)$ and coordinates $x^a$ in the fundamental representation. For traceless generators $K^a{}_b$ and $K^c{}_d$ the invariant metric is $\delta^a_d \delta^c_b - \tfrac1n \delta^a_b\delta^c_d$ and $(\lambda,\lambda)=1-\tfrac1n$ for the fundamental representation. Evaluating~\eqref{eq:GenLie} on a vector with components $V^a$ then leads to
\begin{align}
\mathcal{L}_\xi V^a = \xi^b\partial_b V^a - V^b \partial_b \xi^a\,,
\end{align}
the usual Lie derivative for $\mf{gl}(n)=\mf{sl}(n)\oplus\reals$. Moreover, the section constraint $Y\ket{p}\otimes\ket{q}=0$ becomes trivial in this case so that all coordinates $x^a$ can be used at the same time.

The construction given here agrees with the one in \cite{Palmkvist:2015dea},
where $\mathfrak{g}$ is extended to a Borcherds superalgebra $\mathscr{B}$.
The Cartan matrix $A_{ij}$ of $\mathfrak{g}$ ($i,j=1,2,\ldots,r$, where $r$ is the rank of $\mathfrak{g}$)
is then extended to a Cartan matrix $B_{IJ}$ of $\mathscr{B}$ ($I,J=0,1,\ldots,r$),
such that
\begin{align}
B_{00}&=0\,, & B_{ij}&=A_{ij}\,,& B_{0i}&=B_{i0}=-(\lambda,\alpha_i)\,.
\end{align}
We assume both $A$ and $B$ to be non-degenerate, which implies $(B^{-1})_{00}\neq 0$, although the
construction can be generalised to arbitrary Kac--Moody algebras $\fg$.
In the notation used here, the general expression for $Y$ that follows from the construction in \cite{Palmkvist:2015dea} is then
\begin{align}
\sigma Y&= -\,\eta_{AB}T^A \otimes T^B - \bigg(1+\frac1{(B^{-1}){}_{00}}\bigg)
+\sigma\,. \label{Y-expression1}
\end{align}
Since
\begin{align}
A_{ki}(B^{-1})_{i0} = B_{ki}(B^{-1})_{i0} = - B_{k0}(B^{-1})_{00}\,,
\end{align}
the coefficients of the weight $\lambda$ in the basis of simple roots $\alpha_i$ of $\mathfrak{g}$ are given by
\begin{align}
\lambda = -(A^{-1})_{ij}B_{j0}\alpha_i = \frac{(B^{-1})_{i0}}{(B^{-1})_{00}} \alpha_i\,,
\end{align}
and its length squared by
\begin{align}
\big((B^{-1})_{00}\big)^2(\lambda,\lambda)&=(B^{-1})_{0i}B_{ij}(B^{-1})_{j0}\nn\\
&=-(B^{-1})_{0i}B_{i0}(B^{-1})_{00}\nn\\
&=-(B^{-1})_{0I}B_{I0}(B^{-1})_{00}=-(B^{-1})_{00}\,,
\end{align}
from which it follows that (\ref{Y-expression1}) can be rewritten as (\ref{YTevvvvnsor}).

In terms of the present work, and $\fe_9$, the $Y$ tensor (\ref{E9YTensor})
is already manifestly of the form (\ref{YTevvvvnsor}) with
$(\lambda,\lambda)=0$.
It follows from the presentation above that 
a representative of solutions to the strong
section condition is spanned by $\bra0$ and a subspace
representing the M-theory or type IIB branch of the E${}_8$ strong
section condition. The procedure is general and gives a recipe for such an
``oxidisation'' procedure, which can be continued through a series of
duality groups $X_n$ with decreasing rank by sequentially removing nodes
of the Dynkin diagram corresponding to the coordinate module, with
highest weight $\ket\lambda$, each time expressing a representative of
the solutions of the
strong section constraint for $X_n$ as the linear subspace spanned by $\lambda$
and a section for $X_{n-1}$. In general $X_{n-1}$ is the Levi stabilizer of the
representative $|\lambda\rangle$, that reduces when $\lambda$ is a fundamental weight to the algebra whose Dynkin diagram is the one of $X_n$
with the node associated to $\lambda$ removed. The sequence is uniquely determined provided the module $R(\lambda_n)_n$ is irreducible for all $n$, but this is generally not the case. Whenever the module reduces to several irreducible components, there are as many ``oxidised'' algebras $X_{n-1}$ as there are irreducible components. The 
``oxidisation'' procedure therefore generally gives rise to a tree
rather than a linear sequence. For maximal supersymmetry the module
becomes reducible in $D=9$, giving rise to both the type IIB and the
eleven-dimensional supergravity solution. For half maximal the module
becomes reducible in $D=5$, giving rise to both type IIB on K3 and
heterotic solutions. 

\section{Conclusions\label{ConclusionsSection}}
We have performed the first and critical step towards an exceptional
field theory based on E$_9$ or other affine groups, which consists in
the construction of a closed algebra of gauge transformations. Like in the case of E$_8$, extra local and constrained rotations are part of
the gauge transformations. This is connected to the presence of dual
gravity and other (in the present case an infinite number of) mixed
tensors. These extra transformations are shown to be such that they do
not interfere with the dynamics of the
physical part of a vielbein. The precise covariant form of this
dynamics remains to be constructed. Our construction makes heavy use
of Virasoro generators in order to form and use invariant tensors.
We also provide a generalised Scherk--Schwarz reduction, which shows
that our gauge transformations reduce to the ones expected from
two-dimensional gauged supergravity, and predicted by the tensor
hierarchy algebra. A completely generic form of the $Y$ tensor, and
thereby of candidate generalised diffeomorphisms based on any
Kac--Moody algebra was presented.

One important implication from our construction is that the generalised vielbein should parametrise an element of the coset $G/K(G)$, where the group $G$ is constructed from exponentiation of an extended algebra  $\mathfrak{e}_9 \oplus \mathds{R} \, L_{-1}$ (just like the twist matrix \eqref{Uloop}). In such a construction, the generalised vielbein would include all the fields of the theory, including the scaling factor of the metric in the conformal gauge. It is therefore not clear whether the E$_9$ exceptional field theory can be formulated without resorting to the conformal gauge, such as to be manifestly invariant under both exceptional and ordinary two-dimensional diffeomorphisms. 

The additional gauge transformation involving the tensor $\Sigma$ is
highly degenerate. The parameter $\Sigma$ is a section-constrained element of
$R(\Lambda_0)_{-1} \otimes \overline{R(\Lambda_0)} $, whereas it only
enters the generalised diffeomorphism through its projection to
$\mathfrak{e}_9 \oplus \mathds{R} \, L_{-1}$ defined by $C_{-1}$. The
existence of a Courant algebroid (or generalisation thereof)
underlying the algebra of generalised diffeomorphisms ${\cal
  L}_{\xi,\Sigma}$ remains unclear at the moment. Another
feature of the transformations in their present 
form is that they are non-covariant, {\it i.e.}, it is not possible to
introduce tensors as in \cite{Cederwall:2013naa}.
The situation is in that sense identical to that of the
E$_8$ generalised diffeomorphisms of \cite{Hohm:2014fxa}.
In the E$_8$ case, this
was remedied by the introduction of a non-dynamical background
vielbein and its associated Weitzenb\"ock connection
\cite{Cederwall:2015ica}. The corresponding procedure in the present
case remains an open problem.

Our construction lends strong support to the relevance of the tensor hierarchy
algebra\cite{Palmkvist:2013vya}.
We are necessarily led to a situation where the algebra
consists of $T^A_m$, $\dK$, $L_0$ and $L_{-1}$.  Also the 
embedding tensor 
representation matches the level $-1$ part of the tensor hierarchy algebra.
This is the first
instance where additional elements (in this
case $L_{-1}$) are seen in the algebra, and the lesson should be
important for the continuation to higher exceptional algebras (see
\cite{Bossard:2017wxl}).
In the present work, the well developed representation theory for
affine algebras, relying in particular on the presence of a Virasoro
algebra, was of immense help. If one wants to continue to E$_{10}$ or
E$_{11}$ \cite{West:2001as},
the situation is quite the opposite. Still, level expansions
may be helpful, and the existence of a simple generic form for the
generalised diffeomorphisms looks encouraging.
It would be very interesting to see if a generalised
geometry for E$_{10}$ in some way can make contact with the E$_{10}$
emergent space proposal of \cite{Damour:2002cu}.

\subsection*{Acknowledgements}
We would like to thank Franz Ciceri for comments on the manuscript. 
This work is partially supported by a PHC PROCOPE, projet No 37667ZL,
and by DAAD PPP grant 57316852 (XSUGRA). The work of MC and JP is supported by the Swedish Research Council, project no. 2015-04268. The work of GB was partially supported by the ANR grant Black-dS-String. 

\appendix

\section{Some \texorpdfstring{E$_8$}{E8} representations and tensor products}
\label{app:e8reps}

In this appendix, we collect some useful information about $\fe_8$ representations.
First we list the highest weights of those occurring at low levels in the expansions of the
various $\fe_9$ representations:
\bea
r(0)&=&{\bf1}\;,\nonumber\\
r(\lambda_1)&=&{\bf248}\;,\nonumber\\
r(\lambda_7)&=&{\bf3\,875}\;,\nonumber\\
r(2\lambda_1)&=&{\bf27\,000}\;,\nonumber\\
r(\lambda_2)&=&{\bf30\,380}\;,\nonumber\\
r(\lambda_8)&=&{\bf147\,250}\;,\nonumber\\
r(\lambda_1+\lambda_7)&=&{\bf779\,247}\;,\nonumber\\
r(\lambda_3)&=&{\bf2\,450\,240}\;,\nonumber\\
r(\lambda_1+\lambda_2)&=&{\bf4\,096\,000}\;,\nonumber\\
r(\lambda_6)&=&{\bf6\,696\,000}\;.
\eea

The tensor product of two adjoints gives ${\bf 248} \otimes {\bf 248}
={\bf 1}\oplus{\bf 3\,875}\oplus{\bf 27\,000}\oplus{\bf 248}\oplus{\bf
  30\,380}$. The first 
three are the symmetric product and the last two the
antisymmetric. The projection operators on the five irreducible
representations in the tensor product are given by~\cite{Koepsell:1999uj}
\bea
\mathbb{P}_{({\bf1})}^{MN}{}_{PQ}&=&\frac1{248}\eta^{MN}\eta_{PQ}\komma\nonumber\\
\mathbb{P}_{({\bf3\,875})}^{MN}{}_{PQ}&=&\frac17\delta^{(M}_P\delta^{N)}_Q
    -\frac1{14}f^{A(M}{}_Pf_A{}^{N)}{}_Q
    -\frac1{56}\eta^{MN}\eta_{PQ}\komma\nonumber\\
\mathbb{P}_{({\bf27\,000})}^{MN}{}_{PQ}&=&\frac67\delta^{(M}_P\delta^{N)}_Q
    +\frac1{14}f^{A(M}{}_Pf_A{}^{N)}{}_Q
    +\frac3{217}\eta^{MN}\eta_{PQ}\komma\nonumber\\
\mathbb{P}_{({\bf248})}^{MN}{}_{PQ}&=&-\frac1{60}f_A{}^{MN}f^A{}_{PQ}\komma\nonumber\\
\mathbb{P}_{({\bf30\,380})}^{MN}{}_{PQ}&=&\delta^{MN}_{PQ}
         +\frac1{60}f_A{}^{MN}f^A{}_{PQ}\komma\nonumber\\ 
\label{EEightProjectionOperators}
\eea
where indices are lowered and raised with $\eta_{AB}$ and $\eta^{AB}$. 
The structure constants satisfy the identity \cite{Koepsell:1999uj}
\bea
f^E{}_{AG}f_{BEH}f^{GIC}f_I{}^{HD}
=24\delta_{(A}^C\delta_{B)}^D+12\eta_{AB}\eta^{CD}
-20f^E{}_A{}^Cf_{EB}{}^D+10f^E{}_A{}^Df_{EB}{}^C\;.
\eea

\section{Normalisation of the trace}
\label{FormalTrace}

We would like to define a trace on operators acting on the Hilbert space
(the representation $R(\Lambda_0)$). The idea is that even if
infinities are encountered, they may be consistently renormalised, or
even cancel in final results of calculations. It is included here as a
speculation. If the trace can be defined in a more rigorous way,
it may be useful, since it seems to give correct results at least in
some calculations (see below), but we should stress that we
have not relied on its use in the derivation of any results in the paper.

The relation
of the trace to the quadratic Casimir implies that for some possibly
infinite factor $\cN$, one must have  
\begin{align}
\label{NormalizedTraces} \Tr \, \id =0\ , \qquad \Tr\, L_0 = \cN \ ,
\qquad \Tr\,  T^A_n T^B_m = -\cN  \delta_{m+n,0} \eta^{AB} \ ,  
\end{align}
on the representation space $R(\Lambda_0)$ of the basic module with character 
\begin{align}
(q j(q))^{1/3} = \frac{E_4(q)}{\prod_{n>0} (1-q^n)^8}\;,
\end{align}
where 
\begin{align}
E_4(q) = 1+ 240 \sum_{n>0} \sigma_3(n) q^n = \Theta_{E_8}(q) =
\sum_{Q\in E_8} q^{Q^2/2} \;,
\end{align}
is the theta function of the E$_8$ lattice and the full character is
the partition function of eight free chiral bosons on the E$_8$
torus. The Hilbert space factorises into the momentum component in the
E$_8$ lattice and the oscillator Hilbert space $R(\Lambda_0) = E_8 \otimes
{\mathcal{H}}^{\otimes 8}$, and the action of $L_0$ on $R(\Lambda_0)$ is simply
the tensor product action on E$_8$ and ${\mathcal{H}}^{\otimes 8}$. The naive computation of \eqref{NormalizedTraces} from the Hilbert space trace gives infinite factors for all of them, and one needs to introduce some well chosen insertion to potentially regularise them. One difficulty is to find a regularisation that preserves $E_9$ invariance. We shall simply assume that it exists in the following.

The trace satisfies 
\begin{align}
\Tr_1 ( \overset{1}{X} \sigma_{12} ) = \overset{2}{X}
\; . 
\end{align}
Say that $|J\rangle \langle J|
\in \mathfrak{e}_9$, then one can decompose it in the base $\id,\,
L_0,\, T_n^A$, and one can define a projector using the trace formula  
\begin{align}
 |J\rangle \langle J|  &= \frac{1}{\cN}  \bigl( \Tr\,  |J\rangle
 \langle J| \cdot L_0 +  \Tr\,  L_0 |J\rangle \langle J| \cdot \id -
 \sum_n \eta_{AB}  \Tr\,  T_n^A |J\rangle \langle J| \cdot T^B_{-n}
 \bigr) \nn\\ 
 &=\frac{1}{\cN}  \langle J | C_0 | J \rangle \ . 
 \end{align}
This permits to prove the identity
\begin{align}
\label{cond1} \langle \overset{1}{J} | \sigma_{12} \overset{13}{X}
|\overset{1}{J}  \rangle = \frac{1}{\cN} \Tr_4 \bigg(
\langle\overset{1}{J}  | \Kmat{14}_0 |\overset{1}{J} \rangle
\sigma_{42}  \overset{43}{X} \bigg) = \frac{1}{\cN}   \overset{23}{X}
\langle \overset{1}{J} | \Kmat{12}_0 | \overset{1}{J}  \rangle  
\end{align}
for any operator $X$ acting on the tensor product $R(\Lambda_0) \otimes R(\Lambda_0)$. In
the same way one obtains  
\begin{align}
\label{cond2} \langle \overset{1}{J}  | \sigma_{12}
\overset{23}{X} |\overset{1}{J}  \rangle =  \langle \overset{1}{J}
|\overset{13}{X} \sigma_{12}  |\overset{1}{J}  \rangle= \frac{1}{\cN}
\langle \overset{1}{J} | \Kmat{12}_0 | \overset{1}{J}  \rangle
\overset{23}{X} \ . 
\end{align}
These two identities will be very useful in the following.

Based on this formal trace, an alternative computation of the Scherk--Schwarz Ansatz in the absence of
$L_{-1}$ gauging goes as follows. For a twist matrix $U$ solely in E$_9$, such that
$\varsigma=0$ in~\eqref{eq:Uvir}, one can define the Ansatz in terms of matrices using
the normalised trace  
\begin{equation} 
\frac{1}{\cN}  \mbox{Tr}\,  \mathcal{J}   =  \frac{1}{\cN}  \langle
\Jmat{} | \Jmat{} \rangle \langle \partial_J | =  \langle \partial
\upsilon| \, . 
\end{equation}

The Scherk--Schwarz Ansatz written in the Dirac formalism then takes the form 
\begin{align}
|V\rangle &=   U^{-T} | \underline{V}\rangle\ , \nn\\
| \xi \rangle &=   U^{-T} |\underline{\xi} \rangle\ , \nn\\
|\Sigma\rangle \langle \pi_\Sigma | &= \frac{1}{\cN}  e^{-\upsilon}
 \Umat{2}^{-T} \left( \langle \Jmat{1}|
 \underline{\Kmat{12}}_1 | \Jmat{1}\rangle \otimes
| \overset{2}{\underline{\xi}}\rangle\right) \langle \partial_J | \,, 
\end{align}
and we have
\begin{align}
\mathcal{L}_{\xi,\Sigma} |\overset{3}{V} \rangle &= \langle
\partial_V|\xi\rangle \left(  U^{-T}
|\underline{V}\rangle\right) + \langle \partial_\xi | (C_0-1) \left(
U^{-T} | \underline{\xi}\rangle \otimes |V\rangle\right)
\nn\\ 
&\quad + \frac{1}{\cN} e^{-\upsilon}  \langle
\overset{2}{\partial}{}_J|  \Kmat{23}_{-1} \left( \Umat{2}^{-T}
\langle \Jmat{1} | \underline{\Kmat{12}}_1| \Jmat{1}\rangle \otimes |
\overset{2}{\underline{\xi}}\rangle\right)\otimes
\Umat{3}^{-T} | \overset{3}{\underline{V}}\rangle\nn\\ 
&=-U^{-T} \langle \underline{\partial}_J|\underline{\xi}\rangle
|\Jmat{}\rangle\langle\Jmat{}|\underline{V}\rangle
 -U^{-T} \langle \underline{\partial}\rho |
(\underline{C}_0-1) |\underline{\xi}\rangle \otimes
| \underline{V}\rangle \nn\\
&\quad - U^{-T} \langle
\underline{\partial}_J | (\underline{C}_0-1) \Big(|\Jmat{}\rangle
\otimes |\underline{V}\rangle\Big) \langle \Jmat{} |
\underline{\xi}\rangle
+ \tfrac{1}{\cN}  U^{-T} \langle \Jmat{}| \otimes
\langle \underline{\partial}_J| \underline{\Kmat{23}}_{-1}
\underline{\Kmat{12}}_1 |\Jmat{}\rangle \otimes
|\underline{\xi}\rangle\otimes | \underline{V}\rangle\nn\\ 
&= U^{-T} \bigg[ -\langle
  \underline{\partial}_J |\Jmat{}\rangle \langle\Jmat{} |
  \underline{C}_0 |\underline{\xi} \rangle \otimes
  |\underline{V}\rangle \nn\\ 
&\quad +\langle \Jmat{}|\otimes \langle \partial_J| \bigg(
  -\sigma_{13} + \sigma_{12} (1- \underline{\Kmat{13}}_0
  +\underline{\Kmat{23}}_0) - \tfrac{1}{\cN}
  \underline{\Kmat{23}}_{-1} \underline{\Kmat{12}}_1 
\bigg) |\Jmat{}\rangle \otimes |\underline{\xi}\rangle\otimes
|\underline{V}\rangle\bigg]\,. 
\end{align}

We can now remove all the $\sigma_{12}$ and $\sigma_{13}$ operators
using equation \eqref{cond1} as 
\begin{align}
&\langle \Jmat{}|\otimes \langle \partial_J| \bigg( -\sigma_{13} +
  \sigma_{12} (1-\underline{\Kmat{13}}_0 +\underline{\Kmat{23}}_0)+
  \tfrac{1}{\cN} \underline{\Kmat{23}}_{-1} \underline{\Kmat{12}}_1 
\bigg) |\Jmat{}\rangle \otimes |\underline{\xi}\rangle\otimes
|\underline{V}\rangle\nn\\ 
&=\tfrac{1}{\cN} \langle \Jmat{}|\otimes \langle \partial_J| \bigg(
\underline{\Kmat{23}}_{-1}\underline{\Kmat{12}}_1
-\underline{\Kmat{13}}_0 +
\underline{\Kmat{12}}_0-\lb\underline{\Kmat{23}}_0,\underline{\Kmat{12}}_0\rb
\bigg) |\Jmat{}\rangle \otimes |\underline{\xi}\rangle\otimes
|\underline{V}\rangle\nn\\ 
&=\tfrac{1}{\cN}  \langle \Jmat{}|\otimes \langle \partial_J| \bigg(
\underline{\Kmat{12}}_{1}\underline{\Kmat{23}}_{-1}-\underline{\Kmat{23}}_0
\bigg) |\Jmat{}\rangle \otimes |\underline{\xi}\rangle\otimes
|\underline{V}\rangle \ ,  
\end{align}
where we used \eqref{Commut} in the last step.

The final result is
\begin{align}
U^T \mathcal{L}_{\xi,\Sigma} |V\rangle &= \tfrac{1}{\cN}\bigg[
  \langle \Jmat{}|\otimes \langle \underline{\partial}_J|
  \underline{C}_{1} |\Jmat{}\rangle \bigg] \underline{C}_{-1}
\left(|\underline{\xi}\rangle\otimes
|\underline{V}\rangle\right)\nn\\ 
&\quad  -\bigg[  \langle
  \underline{\partial}_J |\Jmat{}\rangle \langle\Jmat{} |
  +\tfrac{1}{\cN} \langle \Jmat{} | \Jmat{} \rangle \langle
  \underline{\partial}_J|\bigg] \underline{C}_0 \left(
|\underline{\xi} \rangle \otimes |\underline{V}\rangle \right)\nn\\ 
&= \langle \underline{\theta}| \underline{C}{}_{-1} |
\underline{\xi}\rangle\otimes |\underline{V}\rangle +  \langle
\underline{\vartheta} | \underline{C}{}_0| \underline{\xi}\rangle
\otimes |\underline{V}\rangle\,,
\end{align}
corresponding to the ordinary gauging and the $L_0$-gauging.

\bibliographystyle{utphys}

\providecommand{\href}[2]{#2}\begingroup\raggedright\endgroup

\end{document}